\documentclass[vecphys]{svmult}

\usepackage{makeidx}         
\usepackage{graphicx}        
\usepackage{multicol}        

\makeindex             
\newcommand{\be}{\begin{equation}}
\newcommand{\ee}{\end{equation}}
\def\bea{\begin{eqnarray}}
\def\eea{\end{eqnarray}}
\def\cn{{\cal N}}
\def\IP{\relax{\rm I\kern-.18em P}}
\usepackage{amssymb,amsmath}

\begin{document}

\title*{ON INFLATION IN STRING THEORY}
\author{Renata Kallosh}
\institute{Department of Physics, Stanford University, Stanford, CA 94305
\texttt{kallosh@stanford.edu}}

%
%
\maketitle

\

 In this talk we describe  recent progress in construction of inflationary
models in the context of string theory with flux compactification and moduli
stabilization. We also discuss a possibility to test string theory by
cosmological observations\footnote{Based on a talk given at the 22nd IAP Colloquium, ``Inflation +25'', Paris, June 2006.}.

\

\section{Introduction}
\label{sec:1}

In this talk we will discuss problems of string theory in  explaining the cosmological  observations and some  recent progress in construction of 
inflationary models in the context of  flux compactification and moduli stabilization\footnote{Recent reviews on flux compactification and moduli stabilization can be found in \cite{Grana:2005jc,Douglas:2006es,Blumenhagen:2006ci}  and on inflation in string theory in \cite{HenryTye:2006uv,Cline:2006hu}.}. In view of the available precision observational data supporting inflationary cosmology, as well as the new data expected to come  in few years from now, we will  also discus possibilities to test string theory by
cosmological observations.

It is important to find out how the observational cosmology can probe string theory, since our universe is an ultimate test of fundamental physics. High-energy accelerators will probe the scale of energies way below GUT scales. Cosmology and astrophysics are the major sources of data in the gravitational sector of the fundamental physics (above GUT, near Planck scale). 

One can argue that M/string theory is  fundamental: it has sectors with  perturbatively finite quantum gravity. It includes 
supersymmetry and supergravity and has a potential to describe the standard model particle physics and beyond it. It selects d=10 critical string theory and d=11 M-theory. These two dimensions are  
 also maximal dimensions for supergravity, d=10 for chiral supergravity and d=11 for the non-chiral one. 
These theories are almost unique. And in any case, it is
the best and most advanced theory beyond standard model that we have now. But does it have any falsifiable predictions for cosmology? 

To face the observational cosmology one usually assumes the existence of some effective  four-dimensional  ${\cal N}=1$  supergravity based on flux compactification and moduli stabilization, derivable from superstring theory. In  this context string theory has already provided a possible explanation of the dark energy of the universe via an effective cosmological constant
of the metastable de Sitter vacua, \cite{Kachru:2003aw}. The most recent analysis of the data on dark energy  \cite{Wright:2007vr}  confirms the consistency of the cosmological $\Lambda$CDM concordance model with the simplest form of dark energy, the cosmological constant.  The data on dark energy in \cite{Wright:2007vr} are taken from supernovae, gamma ray bursts, acoustic oscillations, nucleosynthesis, large scale structure, and the Hubble constant.
The idea of the landscape of string vacua \cite{Bousso:2000xa,Kachru:2003aw,Susskind:2003kw}  supports the  possibility of an anthropic explanation of the observable value of the cosmological constant.

Several models of inflation were derived since 2003 in the compactified string theory with the KKLT scenario of  moduli stabilization \cite{Kachru:2003aw}. Prior to this recent progress,  string theory had a major problem of runaway moduli. Many interesting ideas were suggested, but the runaway moduli did not allow to have any type of internally consistent cosmology,  see for example  \cite{Quevedo:2002xw,Banks:1995dp,Binetruy:1986ss}.

The first string inflation model based on the KKLT construction, with all moduli stabilized at the exit from inflation, is the  brane-anti-brane annihilation scenario in the warped geometry, the KKLMMT model \cite{Kachru:2003sx}. This model belongs to a general class of brane inflation models \cite{Dvali:1998pa,Quevedo:2002xw} where the inflaton field, whose evolution drives inflation, is associated with the relative position of branes in the compactified space. Another class of string inflation models, which we will discuss later, modular inflation, does not consider brane dynamics. It assumes that the inflaton is one of the many moduli fields present in the KKLT construction. 

In the new models the inflaton field is the only field (or some combination of fields) which is not stabilized before the exit from inflation. Each of these models relies on particular assumptions. Some of these models have clear predictions for observables and are therefore falsifiable by data. Some other models are more speculative and need more work before they can give definite predictions. There is an issue of fine-tuning and the problem with identifying stringy quantum corrections, which require much better understanding. 

The future developments in string cosmology and our attitude  towards various   models of inflation may depend strongly on several crucial pieces of information, which may become available during the next few years. Here is the list of most important observables, which may shift the interest from one class of models of  inflation to another\footnote{See also the contribution of D. Lyth in the Proceedings of this conference with the review of models of inflation  constructed during the last 25 years.}.

\vskip 3mm

A) A precision measurements of the tilt of the spectrum of scalar perturbations, $n_s$, which provides the measure of the violation of the scale invariance,  $n_s-1$. 

\

The current value of  $n_s$, which takes into account the  WMAP3 results, is  close to $0.95$, if one ignores a possible contribution of the    gravitational waves from inflation and from cosmic strings \cite{Spergel:2006hy,Tegmark:2006az}. This value is  below the WMAP1 value, which was about $0.98$. As an example of potential importance of future clarification of the value of spectral index we may refer to D-term inflation in supergravity \cite{Binetruy:1996xj,Halyo:1996pp} and their string theory version,  D3/D7 brane inflation \cite{Dasgupta:2002ew,Koyama:2003yc}. These models  naturally have $n_s=0.98$ and no gravitational waves. This was the perfect value for WMAP1, but it may be on a high side for WMAP3. On the other hand some models of modular inflation in string theory \cite{Blanco-Pillado:2004ns, Lalak:2005hr,Conlon:2005jm,Blanco-Pillado:2006he,Bond:2006nc}  with $n_s \sim 0.95-0.96$ originally where looking not so good, but became much more attractive after WMAP3.
New data on $n_s$ will  provide a powerful tool for selection  of valid models on inflation.
        
\

B) A possible discovery of primordial gravitational waves  from inflation, 
i.e. the measurement of the tensor to scalar ratio $r= T/S$. 

\

The current limit is given by $r<0.3$. A new series of observations may possibly test the models with $r \gtrsim 10^{{-3}}$.  The simplest model of chaotic inflation \cite{Linde:1983gd} with $m^{2}\phi^2$ potential predicts $r\approx 0.15$, with analogous prediction for the chaotic inflation in supergravity \cite{Kawasaki:2000yn}.  This level is expected to be attained during the next few years, particularly at Planck and specialized polarization experiments, like BICEP (up $r \gtrsim 5\cdot 10^{{-2}}$), Spider (up $r \gtrsim 10^{{-2}}$) and others, perhaps all the way to $r>10^{-3}$. It has been clarified recently in \cite{Baumann:2006cd} and in  \cite{Bean:2007hc} that all known models of brane inflation, including the DBI inflation model \cite{Silverstein:2003hf,Alishahiha:2004eh}, do no lead to a prediction of observable $r$ 
\footnote{We are grateful to D. Baumann, R. Bean, D. Lyth, L. McAllister and H. Tye for the discussion of this issue.}. The hope remains that the new brane inflation models with tensors may be constructed.

The model of assisted inflation \cite{Liddle:1998jc,Kanti:1999ie} and related to it proposal of  N-flation model of string theory \cite{Dimopoulos:2005ac,Easther:2005zr} are basically reducible to a chaotic inflation with the corresponding level of observable gravity waves.  We will discuss below to which extent such models can be actually derived from string theory. Other models of string  inflation typically predict $r < 10^{-3}$,  which would make  tensor perturbations almost impossible  to detect.  

The discovery/non-discovery of tensor fluctuations would be crucial for the selection of inflationary models. A discovery of gravitational waves with $r \sim 10^{-1}-10^{-3}$  would make it very important to understand whether inflationary models predicting large $r$  can be  derived from string theory. It would  eliminate a majority of other models of brane inflation and/or modular inflation, which  predict a non-detectable level of gravitational waves. 

\

C) A possible discovery of cosmic strings produced by the end of inflation.\footnote{A detailed discussion of this topic can be found in the contribution of M. Sakellariadou in the Proceedings of this conference, \cite{Sakellariadou:2007bv}.}

\

It has been recognized recently that the discovery of cosmic strings produced by the end of inflation may be one of the most compelling potential observational   windows into physics at the string scale   \cite{Kachru:2003sx,Copeland:2003bj,Polchinski:2004ia,HenryTye:2006uv}. The main point here is  that the current CMB experimental bound\footnote{A stronger bound $G\mu\leq 1.5\times 10 ^{-8}$ has been recently claimed from the Parkes Pulsar Timing Array project \cite{Jenet:2006sv}. The full project is expected to be able either to detect gravity waves from the cosmic strings or reduce the limit to $G\mu\leq 5\times 10 ^{-12}$.} on the tension of cosmic strings, $G\mu\leq 2\times 10 ^{-7}$ \cite{Pogosian:2006hg,Seljak:2006hi}  is difficult to achieve  generically  and simultaneously predict the existence of light cosmic strings  satisfying the bound. If however, the signal from such light cosmic strings will be discovered via B-polarization due to vector modes, the preferred class of models of inflation in string theory may be associated with warped throat geometry as  in various versions of the KKLMMT model   \cite{Kachru:2003sx,HenryTye:2006uv,Cline:2006hu}. This is the basic class of models with a natural suppression mechanism for the tension of cosmic strings due to the position along the warped  throat in the 5th dimension,  which changes the energy scale of the four-dimensional physics.  Another mechanism of production of cosmic strings, satisfying the observational bound on the tension, has been suggested  in strongly coupled heterotic M-theory \cite{Becker:2005pv}.

Other observables, e. g.  the non-gaussianity, may also become important in  future, see \cite{Creminelli:2006rz} for the most recent discussion of this issue.

In Sec.  2 of this paper we will discuss  relations between cosmology and particle physics phenomenology and in Sec. 3 we will discuss an impact of string theory/supergravity  on some issues in cosmology. In particular, we will describe some interesting cosmological models  based on ${\cal N}=1$ supergravity models,  which  have not yet been implemented in string theory. In Sec. 4 we will discuss some brane inflation models. In Sec. 5 the set of modular inflation models, which do not require the presence of branes, is presented. In Sec. 7 we discuss the N-flation/assisted inflation models. In Discussion we focus on possible fundamental reasons for the flatness of the inflaton potentials from the perspective of string theory.

\section{Cosmology and particle physics phenomenology}

For a long time we did not have any string theory interpretation of the acceleration of the universe. This problem was resolved in 2003 with the invention of the  KKLT scenario  \cite{Kachru:2003aw} and its generalizations. By construction, the moduli are first stabilized in some anti de Sitter space with a negative cosmological constant (CC).  The relevant K\"{a}hler potential and superpotential in the simplest case are
\be
K=-3 \ln (T+\bar T)\ ,  \qquad  W= W_0 + A e^{-a T} \ .
\label{KKLT} \ee
Here $W_0$ in the superpotential originating from fluxes stabilizing the axion-dilaton and complex structure moduli. The exponential term comes from gaugino condensation or wrapped brane instantons.
This scenario requires in addition  some mechanism of uplifting of the AdS vacua to a de Sitter space with a positive CC of the form $\delta V= {C\over (T+\bar T)^n}$. In all known cases this procedure  always leads to metastable de Sitter vacua, see Fig. 1 for the simplest case of the original KKLT model. In addition to the dS minimum at some finite  value of the volume modulus $\sigma= {T+\bar T\over 2}$,  there is always a Dine-Seiberg Minkowski vacuum corresponding to an infinite ten-dimensional space with an infinite volume of the compactified space, $\sigma \rightarrow \infty$. The lifetime of  metastable dS vacua usually is much greater than the lifetime of the universe.

\begin{figure}
\centering
\includegraphics[height=4cm]{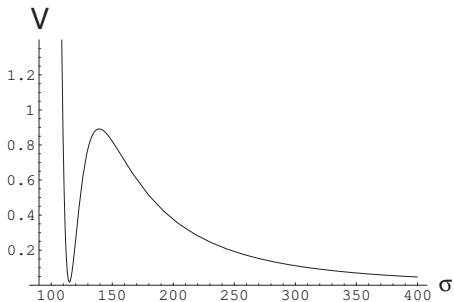}
\caption{KKLT potential as a function of  the volume of extra dimensions $\sigma= {T+\bar T\over 2}$.}
\label{2.ps}       
\end{figure}

There are numerous ways to find flux vacua in string theory, with all possible values of the  cosmological constant.  This is known as the landscape of string vacua \cite{Bousso:2000xa,Kachru:2003aw,Susskind:2003kw}. The concept of the landscape has already changed quite a few settings in particle physics phenomenology. The first and most striking example is that of the split supersymmetry \cite{Arkani-Hamed:2004fb} where the new ideas of supersymmetry breaking where consistently realized without a requirement that supersymmetry has to protect the smallness of the Higgs mass. 

New ideas of particle phenomenology in the context of supergravity and moduli stabilization were developed in \cite{Cascales:2003wn,Camara:2004jj,Lust:2004dn, Choi:2005ge,Conlon:2005ki,Blumenhagen:2006ci,Lebedev:2006qc,Acharya:2007rc,Cremades:2007ig}, leading to a set of new predictions for the spectrum of particles to be detected in the future. 

Recent progress in dS vacuum stabilization in string theory influenced particle phenomenology  by demonstrating that  metastable vacua are quite legitimate. This triggered  a significant new trend   in   supersymmetric model building,   starting with the work \cite{Intriligator:2006dd}.  The long-standing prejudice, that the models of dynamical supersymmetry breaking must have no supersymmetric vacua, is abandoned.   New metastable positive energy vacua with the lifetime longer than the age of the universe were found in supersymmetric gauge models.  
Models with metastable vacua represent an interesting and valid alternative approach to particle phenomenology.


\section{String theory inspired supergravity  models  and cosmology}

Traditionally  cosmological models of inflation use a single scalar field with a canonical kinetic term of the form $ L_{kin}={1\over 2}  \;    \partial_{\mu} \phi \partial_{\nu}  \phi \, g^{\mu \nu}$ with  some particular self-interaction, like ${1\over 2} m^2\phi^2$ or ${1\over 4}\lambda \phi^4$.

In supersymmetric models  one cannot have single scalars fields.
Scalars come in pairs, they are always complex in supersymmetry. For example, axion and dilaton, or axion and radial modulus. Generically, in supergravity and string theory  there is a multi-dimensional moduli space, the scalar fields $\phi^i, \bar \phi^{\bar \imath}$ playing the role of complex coordinates in K\"{a}hler geometry. 
\be
 L_{kin} =   G_{i \bar \imath}(\phi, \bar \phi)  \partial_{\mu} \phi^i \partial_{\nu} \bar \phi^{\bar \imath}\, g^{\mu \nu} \ , \qquad G_{i \bar \imath}\equiv \partial_i \partial_{\bar \imath} K(\phi, \bar \phi) \ .
\label{kin}\ee
The metric in the moduli space $G_{i \bar \imath}(\phi, \bar \phi)$ is derived from the K\"{a}hler potential $K(\phi^i, \bar \phi^{\bar \imath})$, as shown above. An effective inflaton may be a particular direction in moduli space.

 In cases when string theory can be defined by an effective ${\cal N}=1$ supergravity 
the potential is defined by the K\"{a}hler potential and the holomorphic superpotential $W(\phi)$.
\be
V(\phi, \bar \phi)=F_F + V_D=  e^K (|DW|^2- 3 |W|^2)+ V_D \ .
\label{Pot}\ee
Here the total potential consists of the F-term and D-term. The F-term potential $V_F$ depends on K\"{a}hler potential and superpotential whereas the D-term is related to gauge symmetries.

The simple expressions for the  one-field slow-roll inflationary parameters must be generalized in these models, for example,
\be
    \epsilon = {1\over 2} \left (V'\over V\right)^2 \qquad \Rightarrow \qquad   \epsilon = \left( \frac{G^{i\bar \imath} \, \partial_i V
    \partial_{\bar\imath} V}{V^2} \right) \ ,
\ee
where $G^{i\bar\imath}$ is the inverse to $G_{i \bar \imath}$, the Green function in the moduli space, see \cite{Sasaki:1995aw} for more details.

One more comment is due here on the distinctive features of cosmological models in string theory. The scalar fields often have geometrical meaning: distance between branes, size of internal dimensions, size of  supersymmetric cycles on which branes can be wrapped. The axion fields  originate from some  form fields and therefore they are paired into a complex field with particular moduli. In simple examples, which often appear in string theory, 
\be
K= - c \ln (\Phi+\bar \Phi)
\label{example}\ee
with some constant $c$. The origin of the logarithm and shift symmetry (independence on $\Phi-\bar \Phi$) in (\ref{example}) will be explained in Sec. 4.2.
For the total volume-axion $c=3$, which results in no-scale supergravity.
For the dilaton-axion $c=1$ etc. 
The kinetic term for axion and its partner is given by
\be
L_{kin}= c {\partial_\mu \Phi \partial_\nu \bar \Phi g^{\mu \nu}\over (\Phi+\bar \Phi)^2}= {1\over 2} [(\partial \phi)^2 + e^{-2\sqrt{2\over c}\,\phi}(\partial a)^2] \ .
\ee
Here  we take $\Phi= e^{\sqrt{2\over c}\,\phi} +i \sqrt{2\over c}\,  a$ so that  the modulus field $\phi$ has a canonical kinetic term. However, the axion $a$ is coupled to the modulus $\phi$ and this coupling  cannot be removed unless $\phi$ is fixed to a constant value. Typically it is difficult to separate the evolution of the axion and dilaton fields. Both of them are evolving
and both are stabilized only at the exit from inflation, as we will show later.

\subsection{Supergravity models: examples of chaotic and axion valley inflation}

In the effective ${\cal N}=1$ supergravity any choice of a K\"{a}hler potential and a holomorphic superpotential provides a valid theory. 
However, at present only specific versions of the effective ${\cal N}=1$ d=4 supergravity have been derived from a consistent string theory. String theory, in principle, offers a better understanding of quantum corrections. In practice, it remains a major challenge to identify the quantum corrections in the context of non-perturbative compactified string theory with fluxes.  As far as we know now, string theory strongly limits the choice of the effective ${\cal N}=1$ d=4 supergravity models associated with string theory comparative to generic ${\cal N}=1$ d=4 supergravity.

The idea that the  shift symmetry  may help to protect flat directions of the potential was proposed long time ago \cite{Binetruy:1987xj,Banks:1995dp,Gaillard:1995az}. In  the string inspired models it was mostly in the context of  K\"{a}hler potentials $K= - c \ln (\Phi+\bar \Phi)$, or some generic, non-specified K\"{a}hler potentials. At that time it was not known how to stabilize the runaway moduli in such models. As we will  see,  the simple K\"{a}hler-shift symmetric KKLT model, against the naive shift symmetry expectations,  does not have axionic flat directions near the minimum of the potential. 

To have inflation based on the KKLT models we will need to take a more complicated K\"{a}hler potential together with some additional ingredients, see Section 5 on Modular Inflation.
We present below two examples of  ${\cal N}=1$ supergravity models for inflation (which have not been realized in string theory yet), which may explain flat directions for the inflaton field due to shift symmetry with K\"{a}hler potentials of the form $K= {1\over 2}  (\Phi+\bar \Phi)^2$. It will be explained in this section as well as in  Sec. 4.2 that such K\"{a}hler potentials with the shift symmetry do appear in some versions of string theory. 
Here, after presenting the examples,  we will  identify some remaining problems which appear when one tries to implement the simplest versions of chaotic inflation or natural inflation in string theory.

\subsubsection {\it Steep axions in the KKLT-type models with the K\"{a}hler shift symmetry}

\begin{figure}
\centering
\includegraphics[height=5cm]{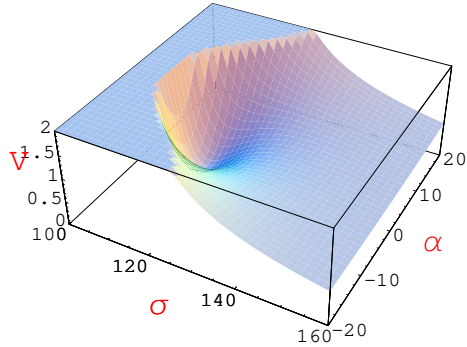}
\caption{The funnel-type potential of the KKLT model depending on the volume $\sigma$ and the axion $\alpha$ from \cite{Kallosh:2004rs}. Fig. 1 in Sec. 2 of this paper gives a slice of this potential in $\sigma$ direction at the minimum for the axion. The potential in the axion direction is as steep as in the volume modulus direction.} \label{KKLTpot.eps}
\end{figure}

The generic property of KKLT models is that the  exponential terms in the superpotential, which stabilize  the volume modulus, simultaneously stabilize the axion field. The masses of these two fields are not much different, as one can easily see in many models, starting from the simplest  KKLT model with  $K=-3 \ln (T+\bar T)$,   $W= W_0 + A e^{-a T}$. The K\"{a}hler potential has the shift symmetry $T\rightarrow T+i\delta$ with real $\delta$.
With $T= \sigma +i\alpha$ the kinetic term for scalars and the potential $V(\sigma, \alpha)$ are:
$$
 L_{kin} = {3\over 4\sigma^2}[(\partial \sigma)^2 +(\partial \alpha)^2] \ ,
$$
$$ 
V= {aA e^{-2a\sigma}\left(A(3+a\sigma)+3 e^{a\sigma}W_0 \cos[a\alpha]\right)\over 6\sigma^2} + {D\over \sigma^3}\ .
$$

Near the minimum of the potential at $\sigma_0, \alpha_0$ the canonical fields are $\sqrt{3\over 2}\, {\sigma\over \sigma_0}$ and $\sqrt{3\over 2}\, {\alpha\over \sigma_0}$. Therefore the curvatures in the axion-volume directions plotted in $\sigma, \alpha$ variables are practically the same for canonically normalized fields. There is no significant flatness in the axion direction comparative to the volume modulus direction near the minimum of the potential.

\subsubsection{\it The   supergravity  version  of chaotic inflation.}

The model proposed in \cite{Kawasaki:2000yn} is based on the K\"{a}hler potential with shift symmetry for the inflaton field 
and with the holomorphic superpotential, which breaks this symmetry,
\be
K= {1\over 2} (\Phi+\bar \Phi)^2+X\bar X \ ,  \qquad W= m\Phi X \ .
\ee
The K\"{a}hler potential does not depend on the inflaton $\varphi= - i(\Phi-\bar \Phi)/\sqrt 2 $, but the superpotential does depend on it. This model has a very steep potential
with respect to all fields except $\varphi$. In the $\varphi$-direction it has a very simple potential 
 ${1\over 2} m^2 \varphi^2$. Thus this model is the supergravity version of the chaotic inflation model for a single scalar field with the potential ${1\over 2} m^2 \varphi^2$.

As of now we do not know whether one can derive  the supergravity chaotic inflation model \cite{Kawasaki:2000yn}   from string theory. 
The reason is that the  K\"{a}hler potentials with shift symmetry for the closed string moduli usually have the form  $K= - c \ln (\Phi+\bar \Phi)$ and therefore would lead to the runaway behavior of the potential of the form $V\sim {1\over (\Phi+\bar \Phi)^c}$. This is very different from $V\sim e^{{1\over 2} (\Phi+\bar \Phi)^2}$, which is an important feature of the model in \cite{Kawasaki:2000yn}. Such type of K\"{a}hler potentials $K= {1\over 2} (\Phi+\bar \Phi)^2$ with shift symmetry have been studied  for the open string moduli \cite{Hsu:2003cy,Firouzjahi:2003zy,Hsu:2004hi} (as we will explain later for D3/D7 brane system). However these fields have restricted range since they correspond to the distance between branes.
It would be very interesting to find a valid regime of  string theory  capable of  reproducing the   supergravity version of the chaotic inflation \cite{Kawasaki:2000yn}, or its generalization, as an effective ${\cal N}=1$ supergravity.
It will be particularly important if both supersymmetry and gravitational waves will be discovered.

\subsubsection{\it The axion valley model (natural inflation in supergravity).}

The natural inflation  PNGB model for the  Pseudo-Nambu-Goldstone Boson  \cite{Freese:1990rb,Adams:1992bn} is based on a potential of the form $\Lambda^4(1\pm \cos (\phi/f)]$ with $f\geq 0.7 M_{Pl}$ and $\Lambda\sim M_{GUT}$.
 In \cite{Adams:1992bn} at attempt was made to derive this potential from string theory with an axion-dilaton field $S$.  The canonical axion field was identified with  $\sim {\rm Im S\over \rm Re S}$. A closely related idea that the axions in string/M-theory may play the role of dark energy  was proposed in \cite{Choi:1999xn}. To obtain  natural inflation or axion dark energy in supergravity it was necessary to stabilize the dilaton, ${\rm Re S}$, and to keep an almost flat potential for the axion, ${\rm Im S}$. Until now, this goal was not achieved. Similarly, no realization of natural inflation or of the axion dark energy was proposed in string theory. 

However, as we are going to show now, it is indeed possible to develop a consistent realization of natural inflation in supergravity. We will consider  the KKLT  model with all fields fixed at their minima, and add to it a field $\Phi$ with a shift-symmetric K\"{a}hler potential and a non-perturbative superpotential  which  breaks the shift symmetry of the K\"{a}hler potential:  
\be
K={1\over 4} (\Phi+\bar \Phi)^2\ ,  \qquad  W= w_0 + B e^{-b \Phi}\ ,
\label{axionvalley}\ee
with\footnote{One can also use two exponents and/or other more complicated version of the model.  }
\be
V_{\Phi}= e^K (|DW|^2- 3 |W|^2)= V_1(x) -V_2(x) \cos (b\beta) \ .
\ee
Here $\Phi= x+i\beta$ and
\be
V_1(x)=e^{x(-2b+x)}B^2(-3+2( x-b)^2+e^{2bx}(-3+2x^2)w_0^2 \ ,
\label{V1}\ee
\be
 V_2=2Be^{bx}w_0(3+2bx-2x^2)  \ . \label{V2}\ee
The presence of the KKLT model is to modify the potential constructed from (\ref{axionvalley}) in two aspects. It  rescales the overall value of the $\Phi$ field potential and adds to it a positive constant. 
The effective uplifting  can make the potential at the minimum of $\Phi$ close to zero (from the positive side). This rescaling can be absorbed by an effective rescaling of $w_0$ and $B$. Thus we have a model with the canonical kinetic term for both $x$ and $\beta$ and the following potential
\be
g^{-1/2} L= {1\over 2}[ (\partial x)^2+ (\partial \beta)^2] - V(x,\beta)\ ,
\label{full}\ee
where the axion valley potential is
\be
V(x,\beta)= V_1(x)- V_2(x) \cos (b\beta)-V_0, \qquad V_0= V_1(x_0)- V_2(x_0)  \cos (b\beta_0) \ , \label{valley}
\ee 
and $x_0,\beta_0$ is the point where the potential has a minimum so that 
the potential vanishes at the minimum. $V_1(x), V_2(x)$ are given in eqs. (\ref{V1}),(\ref{V2}).

\begin{figure}
\centering
\includegraphics[height=4.6cm]{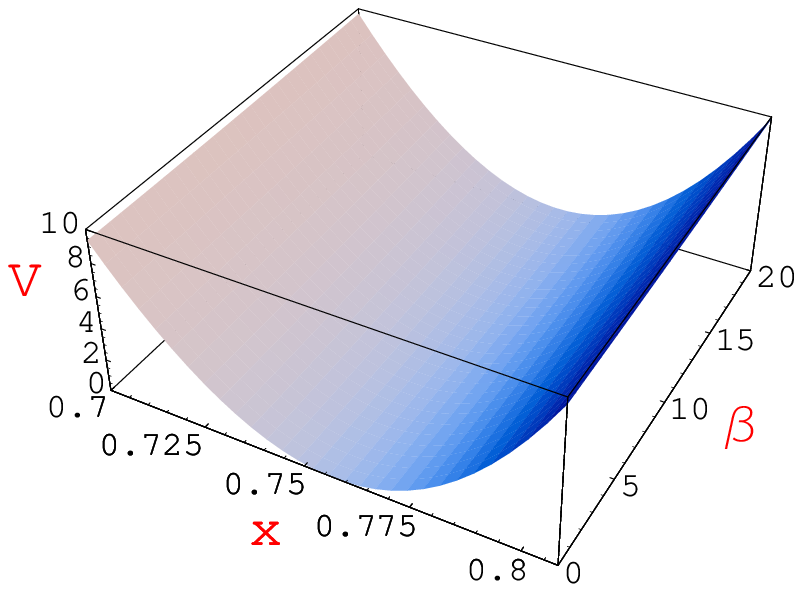} \hskip 0.3cm \includegraphics[height=4.6cm]{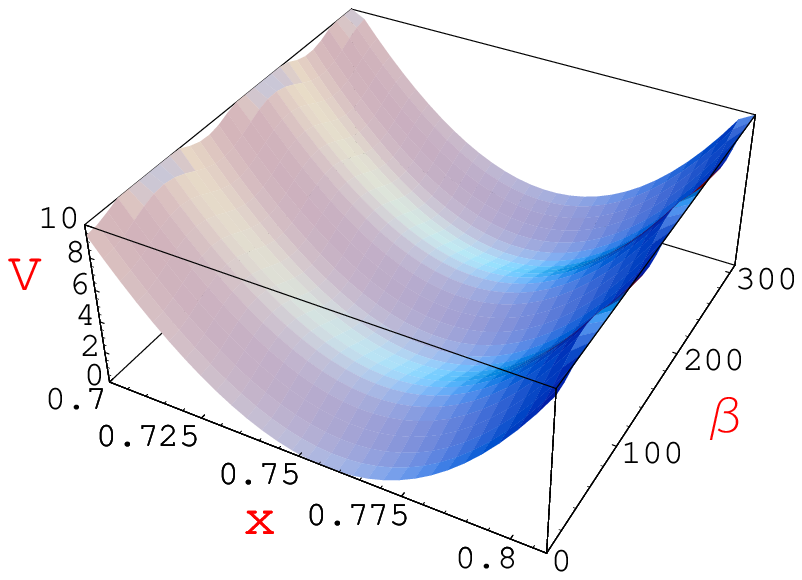}
\caption{Axion valley potential (\ref{full}), (\ref{valley}). On the left figure there is a  view on the axion valley. There is a sharp minimum for $x$ and a very shallow minimum for $\beta$.
The $\beta$-direction is practically flat for $\beta$ from $0$ to $20$ (in Planck units), whereas  in the $x$-direction the potential appreciates significantly when  $x$ changes by $0.1$. On the right figure, the potential is plotted for $\beta$ from $0$ to 300. The plot shows
the periodicity in  the axion variable $\beta$. Both $\beta$ and $x$ have canonical kinetic terms.}
\label{yy2.eps}       
\end{figure}

If the minimum is at $\beta_0=0$ the potential at stabilized $x$ at the point $x_0$ takes the form of the natural PNGB model potential.
\be
V=V_2(x_0)(1 -\cos(b\beta)) \ .
\label{natural}\ee
Our goal is to make the potential in (\ref{full}) for the $x$ field steep and the potential for the $\beta$ field very flat. This is indeed possible, as different from the KKLT model (\ref{KKLT}). In the KKLT model, the potential is equally steep for the volume modulus and the axion  near the minimum of the potential, see  Fig. 2.
Meanwhile, in the axion valley model, e.g. for $B=1$, $b=0.05$, $w_0=10^{-4}$, we find that the potential in the $x$ direction is steep, and we have a nice nearly flat valley for the axion $\beta$, which may play the role of the inflaton field, see Fig. 3 where the potential is also multiplied by $10^3$.

To make this model compatible with the WMAP3 data, we may put the system to the minimum in $x$ at $x=x_0$ and use the values for the parameters suggested in \cite{Savage:2006tr} for the potential  $V=\Lambda^4(1-\cos(\phi/f)$. We need  $V_2(x_0)= \Lambda^4$ with $\Lambda$ at the GUT scale and our parameter $1/b$ corresponds to $\sqrt {8\pi}f$ in \cite{Savage:2006tr}. We have to take into account that in supergravity setting we are working in units where $M= {M_{Pl}/\sqrt {8\pi}}=1$, $M_{Pl}=1.22*10^{19}$GeV. 

There are two limiting cases to consider. In the first case, $0.7 M_{Pl} \leq f\ll 5 M_{Pl}$ ($0.04\ll b\leq 0.28$),  inflation takes place near the maximum of the potential, as in the  new inflation scenario.   In the second case,  $f\geq 5 M_{Pl}$ ($b\leq 0.04$), the potential is very flat at the minimum and the model is close to the simplest chaotic inflation scenario with a quadratic potential.  In this regime, for $x_0<1$,   the COBE/WMAP normalization of inflationary perturbations implies that $ w_0 B\, b^2 \sim 1.5\times 10^{-12}$.  
Clearly, such parameters are possible from the point of view of supergravity, particularly with an account of the rescaling mentioned above when combining this model with the KKLT potential for uplifting.

 To the best of our knowledge, the  axion valley model (\ref{axionvalley}) proposed above provides the first explicit realization of the natural inflation in supergravity. It does realize the standard lore that the shift symmetry of the K\"{a}hler potential may protect a nearly flat axion potential. It gives a simple  example of such a model, where  the partner of the axion is stabilized and the total potential has a stable dS minimum.

This model,  may become useful, in principle, if dark energy in future experiments will  prove to be different from the cosmological constant.
It has been shown recently \cite{Brax:2006np} that scalar moduli as quintessence in supergravity either behave as a pure cosmological constant, or violate constraints   coming from the fifth force experiments, such as Cassini spacecraft experiment. This conclusion was reached for the models with $K=-c\ln(T+\bar T)$, and was valid only for the ``radial'' component of the field, $T+\bar T$. 
The axion moduli like $T-\bar T$ or $\Phi-\bar \Phi$ in the axion valley model (\ref{axionvalley}), if used as quintessence, seem to avoid the fifth force constraint. On the other hand, for such models  a double fine-tuning of $V_1(x_0)$ and $ V_2(x_0)$ with accuracy of $10^{-120}$ is required. This is the usual disadvantage of quintessence models as compared to the simple cosmological constant model. We remind the readers that for the cosmological constant one needs a single fine-tuning, which,  hopefully,  can  be addressed by anthropic considerations in the framework of the stringy landscape.

So far this  model has not been derived from the string theory, but it is available in supergravity. 
The main reason why it is not easy to get an axion valley from string theory is that  one has to justify the K\"ahler potential in Eq. (\ref{axionvalley}). This type of the K\"ahler was identified, e.g., in D3/D7 model via an expansion of the logarithmic potential in Eq. (\ref{expand}). However, in this particular situation it is not easy to argue that the superpotential in Eq. (\ref{axionvalley}) can be used. In the same model with the  $K3\times {T^2\over \mathbb{Z}_2}$ compactification one can use the K\"ahler-Hodge manifold   $S0(2,n)\over SO(2)\times SO(n)$ with the K\"ahler potential
$K=\ln[(x_0+\bar x_0)^2-\sum_{i=1}^{i=n}(x_i+\bar x_i)^2]$\,  \cite{Angelantonj:2003zx}. This K\"ahler potential  is perfectly suitable for our purpose after expansion near the minimum for $x_0+\bar x_0$. However, in this situation the exponential terms in the superpotential  originate from the instantons and not from the gaugino condensation 
\cite{Kallosh:2005gs,Saulina:2005ve,Aspinwall:2005ad}. They  have the form $e^{-2\pi n x_i}$ with integer $n$ which will not result in a flat axion direction as in Fig. 3. It is difficult  to get the small factor in the exponent since for the quaternions $x_i$ there is no gaugino condensation which would give $e^{-{2\pi\over N}  x_i}$  where $N$ is the rank of the gauge group and can be a large number.  It may still be possible in  future to find an adequate model in string theory with a simple axion-type evolution in the spirit of the axion-valley model. One may try to use various studies of axions in string theory in \cite{Banks:1996ea,Banks:2002sd,Banks:2003sx,Svrcek:2006yi,Choi:2006za} towards cosmology of string theoretic axions and stringy axion-type inflation with stabilization of all moduli at the exit from inflation.

Thus, as of now,   the supergravity version of chaotic inflation and the axion valley inflation  models have not been implemented in string theory.
In what follows we will  describe some interesting models of inflation in string theory and comment on their attractive features and problems.

\section{Brane inflation in string theory }

\subsection{KKLMMT-like models of inflation in  the warped throat geometry}

The first   inflationary scenario in string theory with compactification of extra dimensions and  stabilized moduli was proposed in the KKLMMT paper \cite{Kachru:2003sx}. This model is a stringy development of the previously introduced concept of brane inflation \cite{Dvali:1998pa,Quevedo:2002xw}. It is based on the KKLT mechanism of moduli stabilization, so that at the exit from inflation all moduli are stabilized. Inflation occurs as in the hybrid inflation scenario \cite{Linde:1993cn}, the distance between mobile $D3$ and ${\overline{D3}}$ branes plays the role of the inflaton field. Hybrid inflation ends due to the brane-antibrane annihilation. Both inflation and the brane-antibrane annihilation take place in the warped throat geometry \cite{Klebanov:2000hb,Giddings:2001yu}. This model without fine-tuning has a so-called $\eta$-problem \cite{Copeland:1994vg}: the inflaton mass squared is large, $m^2\sim H^2$, and inflation is short. In effective supergravity this can be seen as coming from the K\"{a}hler  potential of the form
\be
K= -3 \ln \left [(T+\bar T)-\Phi \bar \Phi\right ] \ ,
\label{can}\ee
where the distance between $D3$ and ${\overline{D3}}$ branes is related to $\Phi$ and $(T+\bar T)$ is related to the volume of the compactification. At some fixed value of   $T+\bar T$ given by $(T+\bar T)_0$ there is a standard K\"{a}hler  potential $K=  \phi {\bar  \phi}$ for the  redefined distance field $ \phi= {\sqrt 3 \Phi\over  \sqrt{(T+\bar T)_0}}$, which comes from the expansion near the minimum of the volume:
\be
K= -3 \ln[(T+\bar T)-\Phi \bar \Phi] \approx -3 \ln([(T+\bar T)_0]+ { \phi {\bar  \phi}}+...
\label{can1}\ee
The inflaton potential for the field $\phi$ has a form
$e^K V_W $ where $V_W=  (|DW|^2- 3 |W|^2)$.  With $e^K=e^{ \phi {\bar  \phi}}$, the $\eta$-problem is due to the $e^K$ part of the potential. Because of this term, the second derivative of the potential  is of order $H^2$ instead of $\sim 10^{-2}H^2$ as required by the flatness of the spectrum of inflationary perturbations. One of the possibilities to improve the situation is to use
a contribution from the superpotential  which cancels the one from $e^K$.
It is possible to avoid the $\eta$-problem in the KKLMMT  class of models via some fine-tuning with account of stringy quantum corrections \cite{Berg:2004ek} leading to  modifications of the superpotential. The most recent detailed investigation of these and relevant issues was performed in \cite{Baumann:2006th,Burgess:2006cb}. In particular, the structure of the superpotential of the form
\be
W=W_0+ A(\phi)e^{-a T}
\ee
was clarified. Here the pre-exponential factor $A(\phi)$ acquires the dependence on the position of the D3 brane in the throat.
However, the complete and predictive phenomenology of these models is still to be worked out. An extended review of this class of models and their generalizations was presented  recently in \cite{HenryTye:2006uv,Cline:2006hu}.

From the perspective of observational  points explained above, this class of models of $D3$ - ${\overline{D3}}$ brane annihilation in the warped throat geometry has the following features: 

A) The value of the tilt of the spectrum $n_s$ is not unambiguously predictable and depends on the details of the model, like the choice of the fine-tuning and/or the choice of the modification of the original model.

B) In all versions of this model the level of primordial gravitational waves from inflation is predicted to be extremely small.

C) Here is the most attractive feature of inflationary models in this class: They  have  $U(1)$ symmetry, and the corresponding cosmic strings can be easily produced by the end of inflation. Such strings can appear in many versions of the hybrid inflation scenario, which may lead to cosmological problems. However, in the KKLMMT model, the tension of these strings can be easily controlled \cite{Kachru:2003sx} by the warp factor in the throat geometry, $e^{- {2\pi K\over 3 g_s M}}$ \cite{Klebanov:2000hb,Giddings:2001yu}. By a proper choice of the integer fluxes $K$ and $M$ one can make the string tension and string  contribution  to the perturbations of density rather small. But under certain conditions these strings may provide a detectable contribution to gravitational waves \cite{Damour:2004kw,Polchinski:2004ia}. They may also contribute to the CMB polarization.
If the B type polarization in CMB will be detected as coming from the vector modes generated from cosmic strings \cite{Seljak:2006hi, Pogosian:2006hg,Jeong:2006pi}, this class of models will become particularly attractive. A discovery of very light cosmic strings  will  force us to look for a better understanding of the  theoretical problems in this class of inflationary models in string theory.

\subsection{On shift symmetry in D3/D7 brane inflation}

 The D3/D7 brane inflation  model  \cite{Dasgupta:2002ew,Koyama:2003yc,Dasgupta:2004dw}  is a stringy version of the D-term inflation in supergravity. This model has number of interesting features. It relies on one of the most theoretically advanced examples of  stabilization of all moduli of M-theory on $K3\times K3$ manifold and type IIB string theory on $K3\times {T^2\over \mathbb{Z}_2}$ orinetifold \cite{Tripathy:2002qw,Angelantonj:2003zx,Kallosh:2005gs,Saulina:2005ve,Aspinwall:2005ad}. The model  has an approximate  shift symmetry \cite{Hsu:2003cy,Firouzjahi:2003zy,Hsu:2004hi}, which results in the required flatness of the inflaton potential.  The original D-term inflation \cite{Binetruy:1996xj,Halyo:1996pp} has a nearly flat potential since the F-term is vanishing. However, in  D3/D7 brane inflation  model the  F-term potential is required for the volume stabilization. Therefore the shift symmetry of the K\"{a}hler  potential is important for the effective supergravity model of inflation in which the volume of the extra dimensions is stabilized.
 The K\"{a}hler  potential is
\be
K= -3\ln \left [(T+\bar T)-{1\over 2}(\Phi +\bar \Phi)^2\right ] \ ,
\label{shift} \ee
where the two-dimensional distance between D3 and D7 branes is related to $(\Phi +\bar \Phi)$ and $(\Phi -\bar \Phi)$ fields and the volume is related to $T+\bar T$.  The K\"{a}hler  potential   does not depend on the $(\Phi -\bar \Phi)$ field, which is therefore an inflaton field \footnote{Inflationary models in supergravity with shift symmetric K\"{a}hler  potential analogous to (\ref{shift}) were studied in \cite{Brax:2005jv}.}.
As before, $(T+\bar T)$ is related to the volume of the compactification. After the volume stabilization at  $T+\bar T = (T+\bar T)_0$, the K\"{a}hler  potential looks as follows:
\be
K= -3\ln[(T+\bar T)-{1\over 2}(\Phi +\bar \Phi)^2] \approx -\ln [(T+\bar T)_0^3]+ {1\over 2}( \phi + {\bar \phi})^2 +...
\label{expand}\ee
This three-level K\"{a}hler  potential does not depend on the inflaton field $( \phi - {\bar \phi})$,  under certain geometric conditions.   This eliminates the $\eta$-problem in the inflaton direction if all quantum corrections are small. An example of such geometric conditions is the case of the $K3\times {T^2\over \mathbb{Z}_2}$ manifold, where the K\"{a}hler  potential of the form closely related to the one in Eq. (\ref{shift}) was derived in \cite{Angelantonj:2003zx} from the ${\cal N}=2$  supergravity structure, where 
\be
K=-\ln \left[i( \bar X^
\Lambda   F_\Lambda -  X^
\Lambda  \bar  F_\Lambda\right] \ .
\label{N2}\ee
Here the  holomorphic prepotential ${\cal F}  $ depends on coordinates $X^\Lambda= (X^0, X^A)$ of the K\"{a}hler manifold and $F_\Lambda\equiv \partial_\Lambda {\cal F}$. This K\"{a}hler  potential is invariant under symplectic transformations associated with duality symmetry in string theory.
\begin{equation}
\left(\begin{array}{c} X \\F \end{array}\right)' =
\left(\begin{array}{cc} A & B  \\
C & D
\end{array}\right)
\left(\begin{array}{c} X \\
F \end{array}\right)
\end{equation}
In the case of   the cubic prepotential, 
\be
{\cal F}={C_{ABC}\over 3!} {X^A X^B X^C\over X^0}\ , \qquad \Lambda =(0, A)\ ,
\ee
the formula (\ref{N2}) always leads to a K\"{a}hler  potential with a manifest shift symmetry
\be
K=-\ln \left [i {C_{ABC}\over 3!}(z-\bar z)^A (z-\bar z)^B (z-\bar z)^C\right ]\ ,
\label{cubic}\ee
where $z^A={X^A\over X^0}$ are the special coordinates of the K\"{a}hler manifold. 
The symplectic invariance of the K\"{a}hler  potential in ${\cal N}=2$  supergravity plays a fundamental role for the attractor mechanism and computation of the entropy of stringy BPS black holes, \cite{Ferrara:1995ih}.

It was explained in \cite{Hsu:2004hi} that the shift symmetry of D3/D7 model is a subgroup of the duality symmetry  of string theory associated with the cubic superpotential ${\cal F}= s(tu-{1\over 2} x^2)$ where $s$ is the axion-dilaton, 
$t$ is the volume of the K3 manifold, $u$ is the complex structure of the ${T^2\over \mathbb{Z}_2}$ manifold and $x$ is the D3 brane position. The formulae (\ref{shift}), (\ref{expand}) for the K\"{a}hler  potential are somewhat simplified version of the exact expression. The purpose of the simplification was to be able to compare it easily with the analogous K\"{a}hler  potential without shift symmetry in (\ref{can}), (\ref{can1}).

An assumption that stringy quantum corrections  do not break badly the tree-level shift symmetry,  which is broken only softly by Coleman-Weinberg type logarithmic corrections to the potential,  replaces the fine-tuning which is often required for the flatness of the inflationary potential in other models.
These logarithmic corrections are present due to Fayet-Iliopoulos terms (magnetic fluxes on D7 brane).
Note that the K\"{a}hler  potential (\ref{shift}), (\ref{expand})  has a shift symmetry 
\be
\phi \rightarrow \phi + i\delta \ , \qquad \delta= \bar \delta
\ee
of the same type as the chaotic inflation supergravity model \cite{Kawasaki:2000yn}. 

Let us compare it to the models of the KKLMMT type. Stringy corrections of the type derived in \cite{Berg:2004ek,Baumann:2006th} must be significant in  models \cite{Kachru:2003sx} to provide the solution of the $\eta$ problem via the superpotential dependence on the inflaton:  These corrections must cancel the $m^2\sim H^2$ contribution from the K\"{a}hler potential. In the case of the  D3/D7 model with the term ${1\over 2}( \phi + {\bar \phi})^2$ in the K\"{a}hler  potential, there is no $\eta$-problem in the $( \phi - {\bar \phi})$ direction if the stringy corrections of the type studied in \cite{Berg:2004ek,Baumann:2006th} are small. An interesting feature of such corrections is that their value  may depend on the stabilized values of other moduli, e.g. some complex structure moduli. The choice of fluxes stabilizing the complex structure moduli may therefore help
to control these corrections and to make them large  or small, depending on the model.

D-term inflation is closely related to theoretical issues of  D-term uplifting \cite{Burgess:2003ic} of AdS vacua to dS vacua in the KKLT type construction. In string theory the relevant  models use fluxes on D7 branes, which upon volume stabilization reduce to Fayet-Iliopoulos (FI) terms in supergravity. In  presence of dark energy/positive CC the significance of  FI terms in supergravity/string cosmology is very important. This led to a recent progress towards a better understanding of the FI terms \cite{Binetruy:2004hh,Elvang:2006jk} in supergravity/string theory. 

Some issues with consistent constructions of D-term uplifting with FI terms    for   de Sitter vacua were only recently clarified, \cite{Choi:2005ge,Villadoro:2005yq, Achucarro:2006zf,Parameswaran:2006jh,Choi:2006bh,Dudas:2006vc,Haack:2006cy,
Cremades:2007ig,Acharya:2007rc}. 
Here one should note that
in string theory there are no constant FI terms, only field-dependent D-terms! They become ``constant'' FI-terms after moduli stabilization in effective supergravity.
The  D-term uplifting and stringy D-terms inflation have been studied  recently  \cite{Brax:2006yq, Rocher:2006nh,Burgess:2006cb} and one can expect more investigations of these issues in future. Also new ideas on F-term uplifting of AdS vacua to dS vacua with positive energy have been proposed recently \cite{Dudas:2006gr,Abe:2006xp,Kallosh:2006dv}. Note that in supergravity it is easy to find AdS vacua with negative CC. The uplifting mechanisms (D-term or F-term) are designed to convert the AdS supergravity vacua into those with positive CC. Therefore the studies of these uplifting mechanisms in a context of effective supergravity may help us with  an answer  to a  {\it profound issue: where the positive energy of the Universe is coming from?} 

To the extent to which D3/D7 brane inflationary model is reducible to the D-term inflation, the situation with three major observational  possibilities is the following:

A) A generic  value of the tilt of the spectrum is $n_s\approx 0.98$, which is a bit higher than the current number relevant to models without gravitational waves from inflation. Therefore here the future precision data on  $n_s$ will be important. It will also require an effort to suppress $n_s$, if necessary,  as in \cite{Bastero-Gil:2006cm,Lin:2006xt}, towards smaller values in a way motivated by the theory.  It would be interesting to find out what kind of string theory quantum corrections  may help in this direction.

B) The model predicts an undetectable level of primordial gravitational waves from inflation.

C) D-term inflation, as well as its stringy version D3/D7 brane inflationary model, have  $U(1)$ symmetry and the corresponding cosmic strings  are generically produced. These strings are heavy unless a special effort is made to get rid of them \cite{Kallosh:2003ux,Kallosh:2001tm,Urrestilla:2004eh,Binetruy:2004hh, Dasgupta:2004dw,Jeannerot:2006jj,Lin:2006xt}.  It does not seem possible to
produce very light cosmic strings which may eventually be detected via the B type polarization in CMB  from the vector modes.

\subsection{DBI inflation}

The idea of DBI inflation \cite{Silverstein:2003hf,Alishahiha:2004eh}  is to consider a relativistic  motion of the mobile D3 brane. The action of the brane is therefore considered in the proper Born-Infeld form without expansion that keeps only the  second derivative terms. There is a limit on the maximal velocity which  is required to make the action consistent. All this leads to an unusual and interesting model of inflation, which, in principle, may predict some non-gaussianity and  significant gravitational waves. The actual predictions seem to depend on specific assumptions on the geometry in which the brane is embedded etc. As an example of such assumptions, we refer to Ref. \cite{Kecskemeti:2006cg}, where it was shown that 
to match the data, D3 must move close to the tip of the warped throat. In models with  60-e-folds in the KS throat \cite{Klebanov:2000hb}, the non-gaussianity was shown to be above the current bounds. However, in other geometries things may be different and have to be studied separately.  One of the interesting features of this model is that  some minimal level of non-gaussianity is always expected: this makes the model falsifiable by the future data\footnote{Most recent analysis of this model in  \cite{Baumann:2006cd} and, particularly in  \cite{Bean:2007hc}, indicates that  the observable $r$ may not be possible in all developed brane inflation models including the DBI model.}.

\subsection{Assisted M5 brane inflation}

We would like to comment here on one more  model of inflation in the heterotic M-theory with multiple moving M5 branes \cite{Becker:2005sg}. This model is an attempt to find a version of assisted inflation \cite{Liddle:1998jc,Kanti:1999ie} in M-theory where a   large amount of branes may help to realize an effectively flat potential to assist the inflation which is not possible for a single brane. The problem  of the heterotic M-theory is that the stabilization of the orbifold length, the volume of compactification and other moduli requires a regime of a strong coupling.
However it is interesting that if we take the prediction  of the phenomenological part of the  model regarding a spectral index we find the following change from WMAP1 to WMAP3 data. For $n_s \approx 0.98$ the number of M5 branes was required for the assistance effect to be about 89. For the new value of the index $n_s\approx 0.95$ one needs 66 branes. This model gives an example of the situation when the new data can be easily accommodated within the same model.

\subsection{A remark on conceptional issues in brane inflation}

All models of brane inflation starting with \cite{Dvali:1998pa} have an interesting and novel feature with regard to previously known models: a possibility of an interpretation of the four dimensional scalars as distance between branes, as excitations of strings stretched between branes, etc. 
This advantage, however, is somewhat difficult to realize in a clear and consistent way in the context of a compactified internal space.

The conceptual problems of brane inflation in general  reflect  the difficulty of describing the action of branes, or a probe brane,  in a background of compactified  internal space. This is a reflection of the major problem of string theory where the description of the open string sector and D-branes is not clearly formulated  in the framework of the closed string theory.

 More work will be required here to clarify the geometry of extra dimensions in presence of moving extended objects and the relevant effective supergravity in four dimensions.

\section{Modular inflation in string theory}

The models of the so-called modular inflation, i.e. inflation with moduli fields corresponding to the closed string sector, are conceptually simpler than the models of brane inflation. Several interesting models belonging to this class have been derived lately; we are going to discuss them below.  

One could wonder why did we start with brane inflation in our investigation of the KKLT-based inflationary models? The answer is simple: we were unable to identify any flat directions suitable for inflation in the original KKLT models. However, perhaps, some simple axion-type inflation models are available  in string theory with stabilized moduli.
In Sec. 3.1 of this paper, we constructed an explicit ${\cal N}=1$ supergravity model which has an axion valley with a stabilized scalar direction and a flat axion direction, see Fig. 3. To the best of our knowledge, this is the first explicit realization of natural inflation in supergravity.
However, so far we were unable to find any such models  in string theory.

We will therefore focus here on the existing modular inflation models derived from string theory: the racetrack inflation \cite{Blanco-Pillado:2004ns,Lalak:2005hr},  better racetrack inflation \cite{Blanco-Pillado:2006he}, large volume K\"{a}hler inflation \cite{Conlon:2005jm} and its generalized version, roulette inflation \cite{Bond:2006nc}. Unlike brane inflation models discussed in the previous section, all of these modular inflation models allow eternal slow-roll inflation.

The spectral index is  $n_s\approx 0.95$ for the racetrack models;  $n_s\approx 0.96$ for the large volume K\"{a}hler inflation. In all cases the amplitude of tensor perturbations is extremely small, and there are no  cosmic strings. If Planck satellite will confirm $n_s$ at this level and will not detect gravity waves and cosmic strings, we will have to pay serious attention to these models.

Their conceptual simplicity comparative to the brane inflation models is due to the fact that one can use the framework of closed string theory only (i.e. without open strings) and relate it to the effective ${\cal N}=1$ d=4 supergravity. One has to deduce the structure of the K\"{a}hler potential and the superpotential and find out the parameters which provide inflation. This is where the fine-tuning enters.

A toy model of racetrack inflation \cite{Blanco-Pillado:2004ns} is based on one complex modulus and has 5 parameters with significant fine-tuning. The gaugino-induced racetrack inflation \cite{Lalak:2005hr} has 2 complex moduli, 9 parameters  with less fine-tuning, and has most features as in \cite{Blanco-Pillado:2004ns}.
More realistic models of string theory compactification, the ``better racetrack models''  \cite{Blanco-Pillado:2006he} start with 2 complex moduli and in case of 5 parameters require a significant fine-tuning. Large volume K\"{a}hler inflation models \cite{Conlon:2005jm,Bond:2006nc} require 3 or more moduli, start with 11 parameters and need less fine-tuning.

\subsection{Racetrack Inflation as Eternal Topological Inflation}\label{topinfl}

Slow-roll inflation is realized if a scalar potential $V(\phi)$
is positive in a region where the following conditions are
satisfied:
\be
\label{slowroll}
\epsilon  \equiv  \frac{1}{2}\ \left( \frac{V'}{V}\right)^2 \ll  \ 1\qquad
\eta  \equiv  1\ \frac{V''}{V}\  \ll  \ 1\ .
\ee
Primes refer to
derivatives with respect to the scalar field, which is assumed to
be canonically normalized. Satisfying these conditions is not 
easy  for typical potentials since the inflationary
region has to be very flat. Furthermore, after finding such a
region we are usually faced with the issue of initial conditions:
Why should the field $\phi$ start in the particular slow-roll
domain?

For the simplest chaotic inflation models of the type of ${m^2\over 2} \phi^2$
this problem can be easily resolved, see e.g. \cite{Linde:2005ht}. 
The problem of initial conditions in the theories where inflation is possible
only at the densities much smaller than the Planck density is much more
complicated; for a possible solution see e.g. \cite{Linde:2004nz}). In any case, one can always argue that even if the probability of
proper initial conditions for inflation is strongly suppressed, the possibility
to have eternal inflation infinitely rewards those domains where inflation
occurs. In other words, one may argue that the problem of initial conditions in the theories where eternal inflation is possible becomes largely irrelevant.

Eternal inflation \cite{Vilenkin:1983xq, Linde:1986fd}
is not an automatic property of all inflationary
models. Many versions of the hybrid inflation scenario, including
some of the versions used recently for the implementation of
inflation in string theory, do not have this important property.
Fortunately,  inflation is eternal in all models where it occurs
near the flat top of the scalar potential. Moreover, in this case eternal inflation occurs even at the classical level,  due to the eternal expansion of topological
defects \cite{Linde:1994hy,Vilenkin:1994pv,Linde:1994wt}.

The racetrack inflation \cite{Blanco-Pillado:2004ns} gives an example of eternal topological inflation
within string theory moduli space, generalizing the KKLT scenario.
The model differs from the simplest  KKLT case only in the form assumed for the
nonperturbative superpotential, which is taken to have the
modified racetrack form 
\be
K=-3 \ln (T+\bar T)\ ,  \qquad  W= W_0 + A \, e^{-a T}+B\,e^{-bT} \ .
\ee
Such superpotential  would be obtained through gaugino condensation in a theory
with a product gauge group.  The constant term $W_0$ results from fluxes and represents the
effective superpotential as a function of all the fields that have
been fixed already, such as the dilaton and complex structure
moduli.
As in the simplest KKLT model, the scalar potential  is a sum of two
parts
$
V= V_F + \delta V \,.
$
The first term comes from the standard $\cn=1$ supergravity
formula for the F-term potential in (\ref{Pot}).
The uplifting potential, $\delta V$, is taken in the form
$
\delta V = {  E\over X^2}$ and $ T\, \equiv X+i Y\,.
\label{sb}
$
The total potential is
\begin{eqnarray}
    V~&= & \frac{E}{X^\alpha}\  +\
    \frac{e^{-aX}}{6X^2}\left [ aA^2\left(aX+3\right)~ e^{-aX}\ +
    3 W_0 aA \cos(aY)\right] + \nonumber \\
    &+&\frac{e^{-bX}}{6X^2}\left [ bB^2\left(bX+3\right)~ e^{-bX}\
    + 3 W_0 bB \cos (bY)\right] + \nonumber \\
    &+&\frac{e^{-(a+b)X}}{6X^2}\left[ AB\left(2abX +3a +3b\right)
    \cos((a-b)Y) \right]
\label{potential1}
\end{eqnarray}
Notice that, to the order that we are working, the K\"ahler
potential depends only on $X$ and not on $Y$. For fields rolling
slowly in the $Y$ direction this feature helps to address the $\eta$
problem of F-term inflation. All dependence on the axion field $Y$ is via $\cos(aY)$, $\cos(bY)$ and $\cos((a-b)Y)$ as the result of the shift symmetry of the K\"ahler potential, which is broken by the axion dependence in the exponents in the superpotential. 

This potential has several de Sitter (or anti-de Sitter) minima,
depending on the values of the parameters $A,a,B,b,W_0, E$. In
general it has a very rich structure, due in part to the
competition of the different periodicities of the $Y$-dependent
terms. In particular, $a-b$ can be very small, as in standard
racetrack models, since we can choose $a=2\pi /M$, $b=2\pi/N$ with
$N\sim M$ and both large integers. Notice that in the limit
$(a-b)\rightarrow 0$ and $W_0 \rightarrow 0$, the $Y$ direction
becomes exactly flat. We can then tune these parameters (and $AB$)
in order to obtain flat regions suitable for inflation.

\begin{figure}[h!]
\centering
\includegraphics[height=3.5 cm]{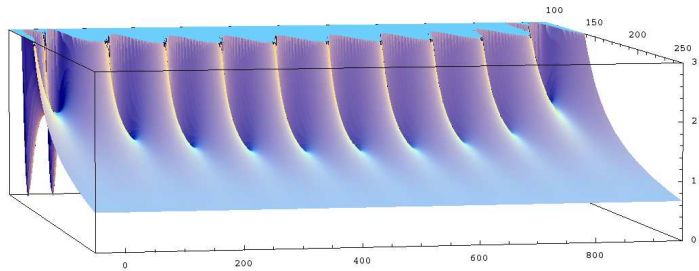}
\caption[fig1] {A plot for the racetrack  potential (rescaled by
$10^{16}$). Inflation begins in a vicinity of any of the saddle points. 
Units are $M_p=1$. As one can see, the potential is periodic in the axion direction, but it is very much different from the potential of natural inflation: there is no axion valley here. \label{longpotpost.eps}}
\end{figure}
The values of the parameters of this potential are:
$
A=\frac{1}{50},  B=-\, \frac{35}{1000}, 
a=\frac{2\pi}{100},  b=\frac{2\pi}{90},
   W_0 = -\frac{1}{25000}\   
$.
Note that
the potential  is periodic with period 900, i. e. there is a set of two
degenerate minima at every  $Y=900\; n$ where $n=0, 1, 2, ...$. etc. as shown in Fig. 4. 

The shape of the potential is very sensitive to the values of
the parameters. 
\begin{figure}[h!]
\centering
\includegraphics[height=5cm]{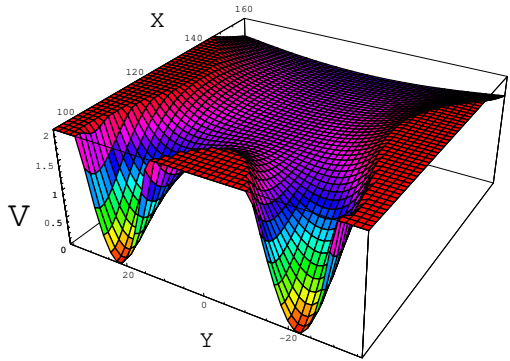}
\caption[fig1] {Plot for a racetrack type potential (rescaled by
$10^{16}$). Inflation begins in a vicinity of  the saddle point
and ends up in one of the two minima, depending on initial conditions. Note that near the minima the potential has a  KKLT type funnel shape where the curvature in volume and axion direction is of the same scale for canonical variables $X/X_{min}$ and $Y/X_{min}$.
\label{Saddle4.eps}}
\end{figure}
Fig.~\ref{Saddle4.eps} illustrate a region of the scalar potential for
which inflation is possible. 
With these values the two minima seen in Fig. ~\ref{Saddle4.eps} occur for
field values
$
X_{min}\ =\ 96.130,$ and $ Y_{min}\ =\ \pm 22.146 \ ,
$
and the inflationary saddle point is at 
$ X_{\rm saddle}=123.22,$ and $ Y_{\rm saddle}=0 \,.
$
It is crucial that this model contains two degenerate minima since
this guarantees the existence of causally disconnected regions of space which
are in different vacua.  These regions necessarily have a domain wall between
them where the field is near the saddle point and thus eternal inflation is
taking place, provided that the slow roll conditions  are
satisfied there.  It is then inevitable to have regions close to the saddle in
which inflation occurs, with a sufficiently large duration to explain our flat
and homogeneous universe.

The racetrack model as well as many other string inflation models has some interesting scaling properties, e. g.
\begin{equation}
\label{rescaling1}
a \rightarrow a/\lambda \,,\hspace{.5cm}
b\rightarrow b/\lambda \,,\hspace{.5cm}
E\rightarrow \lambda^2 E \,,\hspace{.5cm}
\end{equation}
\begin{equation}
\label{rescaling2}
A \rightarrow \lambda^{3/2} A \,,\hspace{.5cm}
B\rightarrow \lambda^{3/2} B \,,\hspace{.5cm}
W_0\rightarrow \lambda^{3/2} W_0 \,.\hspace{.5cm}
\end{equation}
Under all these rescalings the potential does not change under condition that
the  fields also rescale
\begin{equation}
X \rightarrow \lambda X \,,\hspace{.5cm}
Y \rightarrow \lambda Y \,,\hspace{.5cm}
\end{equation}
in which case  the location of the extrema also rescale. One can verify that
the values  of the slow-roll parameters $\epsilon$ and $\eta$ do not change and
also the amplitude of the density perturbations ${\delta \rho\over \rho}$
remains the same. It is important to take into account that the kinetic term in
this model is invariant under the rescaling, which is not  the case for
canonically normalized fields.

Another property of this model is given by the following rescalings
\begin{equation}
\label{rescaling3}
a \rightarrow a/\mu \,,\hspace{.5cm}
b\rightarrow b/\mu \,,\hspace{.5cm}
E\rightarrow  E/\mu \,,\hspace{.5cm}
\end{equation}
\begin{equation}
\label{rescaling4}
V\rightarrow  \mu^{-3} V\,,\hspace{.5cm} X \rightarrow \mu X \,,\hspace{.5cm}
Y \rightarrow \mu Y \,.\hspace{.5cm}
\end{equation}
Under these rescalings the values of the slow-roll parameters $\epsilon$ and
$\eta$ do not change however,  the amplitude of the density perturbations
${\delta \rho\over \rho}$ scales as $\mu^{-3/2}$.

These two types of rescalings allow  to generate many other models from the
known ones, in particular, change the positions of the minima or, if one is
interested in eternal inflation, one can easily change ${\delta \rho\over
\rho}$ keeping the potential flat.

One can study the slow-roll inflation, by examining field motion
near the saddle point which occurs between the two minima
identified above.  At the saddle point the potential has a maximum
in the $Y$ direction and a minimum in the $X$ direction, so the
initial motion of a slowly-rolling scalar field is in
the $Y$ direction.

To compute observable quantities for the CMB we numerically evolve
the scalar field starting close to the saddle point, and let the
fields evolve according to the cosmological evolution equations
for non-canonically normalized scalar fields.
The results of the numerical evolution confirm that the inflaton is primarily in the axionic direction
$Y$ at the very beginning of inflation, as must be the case since $Y$ is the unstable
direction at the saddle point. Eventually both the axion $Y$ and the volume modulus $X$ reach the absolute minimum. The spectral index is found to be $n_s\approx 0.95$ in the COBE region of the spectrum.

\subsection{Better racetrack inflation model}

The KKLT model with stabilization of just one complex K\"ahler  modulus is a simplified toy model of the generic stabilization phenomenon with many moduli. One of the simplest more realistic models of this kind has  two K\"ahler moduli. It
is based on an
explicit compactification of type IIB string theory: the orientifold of 
degree 18 hypersurface 
$\IP^4_{[1,1,1,6,9]}$, an elliptically fibered Calabi-Yau over
$\mathbb{P}^2$. The stabilization of moduli in this model was performed 
in \cite{Denef:2004dm} where it was also shown how D3 instantons
generate a  nonperturbative superpotential, thus providing an explicit
realization of the KKLT scenario.

The model is a Calabi-Yau threefold with the number of K\"ahler
moduli $h^{1,1}=2$ and the number of 
complex structure moduli $h^{2,1}=272$. The 272 parameter prepotential
for this model is not known. 
However, one can restrict ourselves to the slice of the complex structure
moduli space which is fixed under the
 action of the  discrete symmetry $\Gamma\equiv \mathbb{Z}_6 \times
 \mathbb{Z}_{18}$. This allows to 
reduce the   moduli  space of the complex Calabi-Yau structures
to just 2 parameters, since the 
slice is two-dimensional. This restricted model was studied intensely 
in string theory.
  The remaining 
270 moduli are required to vanish to support this symmetry. The
defining equation for the 
Calabi-Yau 2-parameter subspace of the total moduli space is \be
 f=   x_1^{18} + x_2^{18}+ x_3^{18}+ x_4^3 + x_5^2 - 18 \psi x_1 x_2
x_3 x_4 x_5 - 3 \phi x_1^6 x_2^6 x_3^6 \,.
\label{CY}\ee
The axion-dilaton and
all complex structure moduli are stabilized by fluxes. The
remaining 2 K\"ahler moduli are stabilized by  a nonperturbative
superpotential. 
For this model
we identify situations for which a linear combination of the
axionic parts of the two K\"ahler moduli acts as an inflaton. As
in the  previous racetrack scenario, inflation begins at a saddle point of
the scalar potential and proceeds as an eternal topological
inflation.

The K\"ahler geometry of the  two K\"ahler  moduli $h^{1,1}=2$  
was specified in  \cite{Denef:2004dm}.  We denote them 
by $\tau_{1,2}=X_{1,2} +i Y_{1,2}$. These moduli correspond 

geometrically to the complexified volumes of the divisors 
(or four-cycles) $D_4$ and $D_5$, and give rise to the gauge 
couplings for the field theories on the D7 branes which wrap 
these cycles.  For this manifold the K\"ahler potential  is
given by
\be\label{Kahler}
   K = - 2 \ln R = - 2 \ln[\frac{1}{36}\left(({\tau_2+\bar \tau_2})^{3/2} - ({\tau_1+\bar \tau_1})^{3/2}\right)]
   \,,
\ee
where $R$ denotes the volume of the underlying Calabi-Yau space. The flat directions of the potential
are lifted by D3 instantons, which generate the following nonperturbative
superpotential:
\be\label{dosexp}
   W=W_0+A\,e^{-a\tau_1}+B\,e^{-b\tau_2}\,.
\ee
Given these expressions for $K$ and $W$, the  scalar potential takes the following form:
\bea \label{scalarpot2}
    V_F + \delta V && ={216 \over
    ({X_2}^{3/2}-{X_1}^{3/2})^2}\big\{ B^2 b(b {X_2}^2 + 2
    b {X_1}^{3/2} {X_2}^{1/2} + 3 X_2) e^{-2 b X_2}\nonumber \\
    && + A^2 a(3 X_1 + 2 a {X_2}^{3/2} {X_1}^{1/2} + a X_1^2)
    e^{-2 a X_1}\nonumber \\
    &&+3 B b \, W_0 X_2 e^{-b X_2} \cos(b Y_2) + 3 A a\, W_0  X_1
    e^{-a X_1} \cos(a Y_1)\nonumber \\
    &&+3 AB e^{-a X_1 -b X_2} (a X_1
    +b X_2 + 2 a b X_1 X_2) \cos(-a Y_1 + b Y_2))\big\}\nonumber \\
    &&+{D\over \left(X_2^{3/2} - X_1^{3/2}\right)^2}\,
\eea
where the last term is the uplifting term $\delta V$.
Notice that this potential is parity invariant, $(X_i,Y_i) \to
(X_i,-Y_i)$, with $Y_i$ being pseudoscalars. It is also invariant
under the two discrete shifts, $Y_1 \to Y_1 + 2\pi m_1/a$ and $Y_2
\to Y_2 + 2\pi m_2/b$, where the $m_i$ are arbitrary integers.
There is also the approximate $U_R(1)$ $R$-symmetry, $a\delta Y_1 =
b\delta Y_2 = \Delta$, which becomes exact in the limit $W_0 \to
0$. 

We next ask whether slow-roll evolution is possible with this
superpotential and K\"ahler potential. 
Searching the parameter space,  we are able to find choices
for which the scalar potential behaves similarly to
the original racetrack inflation potential.  Starting at the
saddle point,
since only one of the four real directions is unstable, we
have sufficient freedom to make this direction
flat enough to give rise to successful inflation. 

Our goal was to find a set of parameters required for 
inflation with the COBE normalization of power spectrum.
 These examples are not particularly 
easy to find.  The example with $P(k_0) = 4\times 10^{-10}$ and 
$n_{s} = 0.95$, 
has the following parameters:
$
    W_0=  5.22666 \times 10^{{-6}},$
   $ A=0.56,  B=7.46666 \times 10^{{-5}}$,
    $a= 2\pi/40,  b= 2\pi/258, 
 D = 6.21019\times 10^{-9} \ .
$
\begin{figure}
\centering
\includegraphics[height=4.4cm]{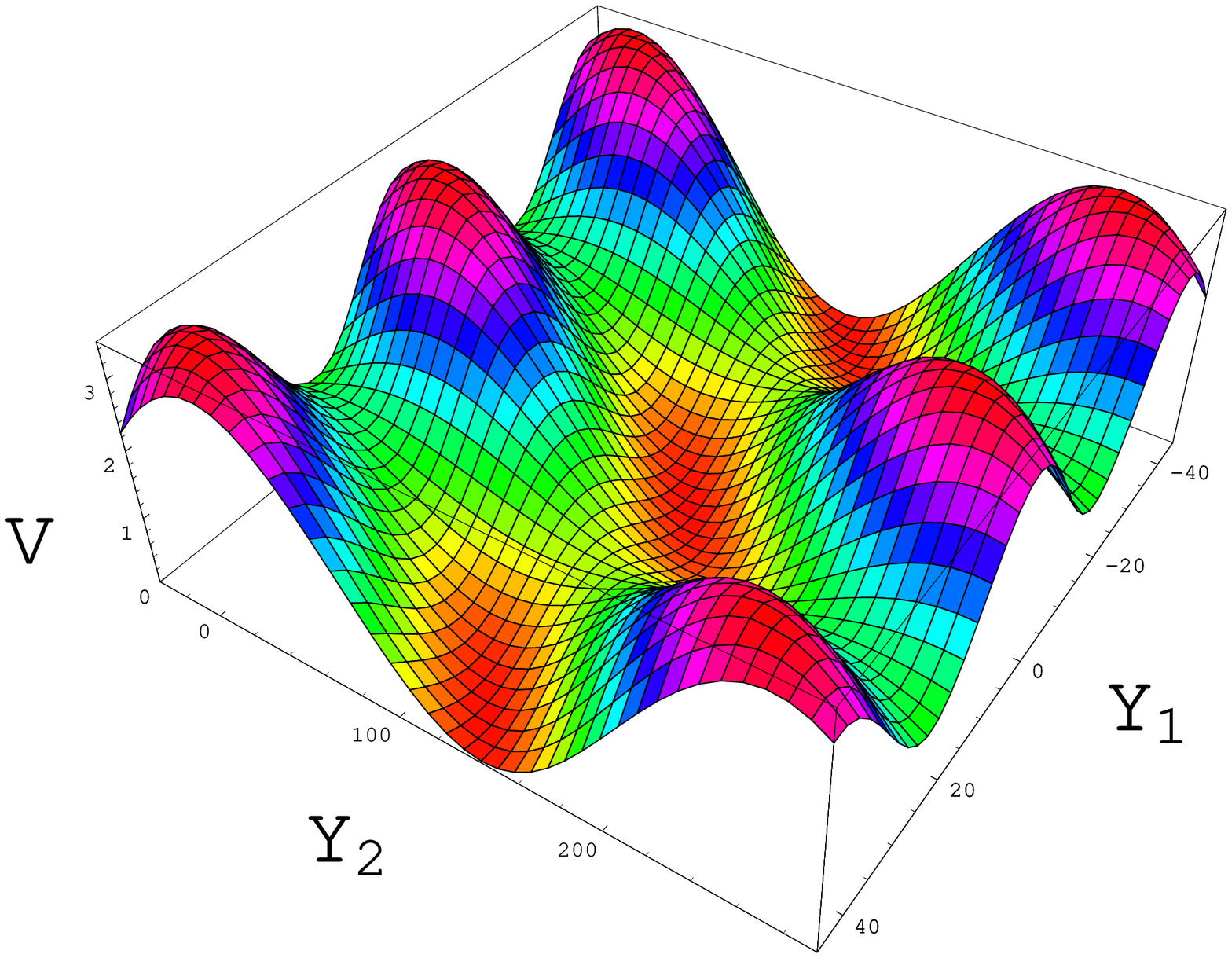}\hskip 1 cm \includegraphics[height=4.4cm]{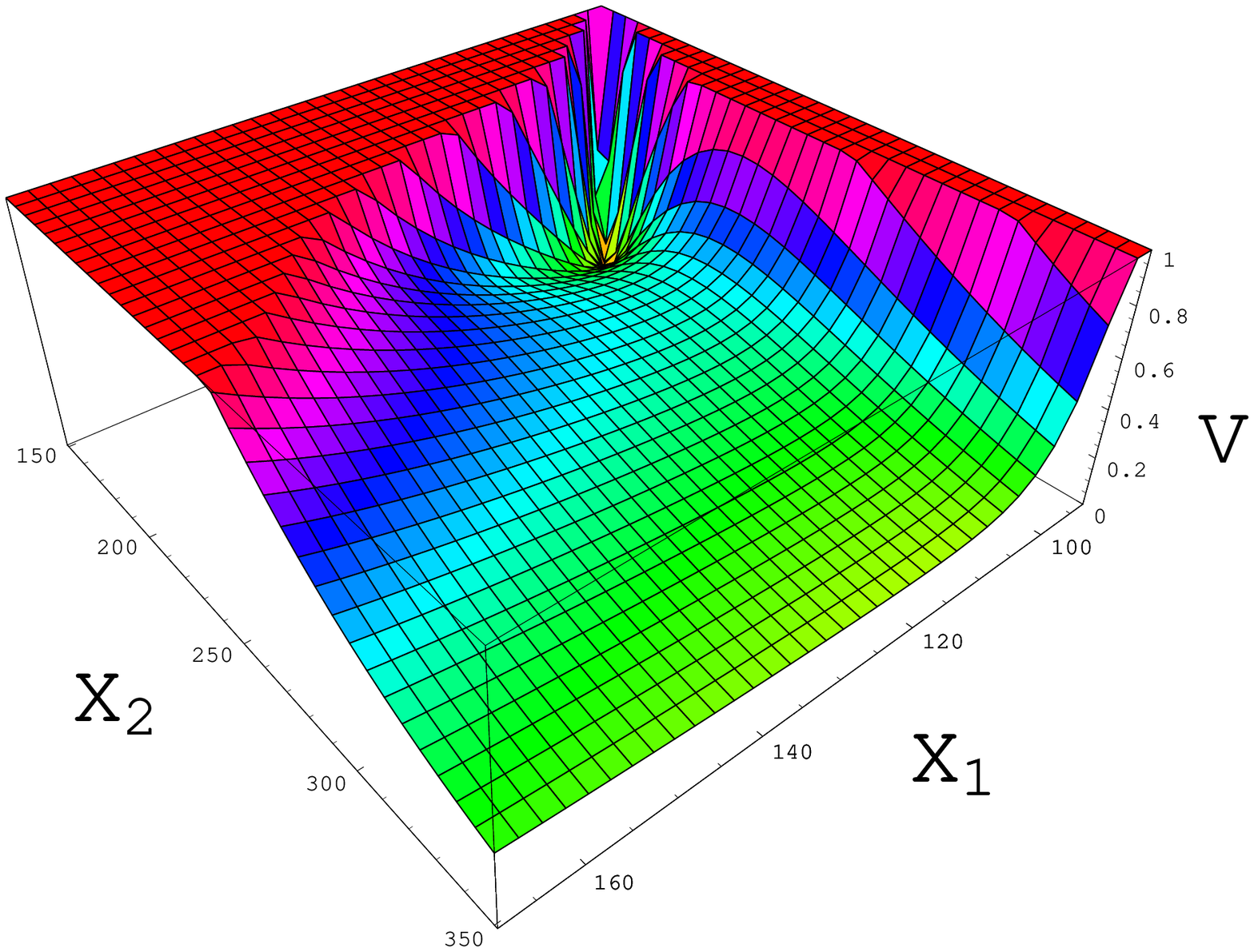}
\caption{On the left there is a potential as a function of the axion variables $Y_1$, $Y_2$
  at the minimum of the radial variables $X_1, X_2$,  in units $10^{{-15}}$ 
of the Planck density. On the right there is a potential as a function of the radial variables $X_1$, $X_2$
  at the minimum of the angular variables $Y_1, Y_2$, in units $10^{{-14}}$ of the Planck density.}
\label{yy.eps}       
\end{figure}
With these choices of the parameters the minimum described above is located at
$
    X_1=98.75839,  X_2=171.06117, Y_1=0,
    Y_2= 129 \,,
$
corresponding to a volume $R=99$ in string units, which is
large enough to trust the effective field theory treatment we use.

It is  difficult to plot the  potential since it is a function 
of 4 variables. Here we will only show the behavior of this potential 
as a function of the axion variables $Y_1$, $Y_2$  at the minimum of 
the radial variables $X_1, X_2$, and the potential as a function of 
the radial variables $X_1$, $X_2$ at the minimum of the angular 
variables $Y_1, Y_2$. Figure \ref{yy.eps} illustrate the behavior of 
the potential near the minimum  of the potential.

We have
checked that the eigenvalues of the Hessian (mass$^2$) matrix are
all positive, verifying that it is indeed a local minimum. The
value of the masses for the moduli at this minimum turn out
to be of order $10^{-6} - 10^{-7}$ in Planck units.
Inflation occurs near the saddle point  located at
$
    X_1=108.96194, X_2=217.68875,
    Y_1=20, Y_2=129 \,.
$
At this point the mass matrix has three positive eigenvalues and
one negative one in the direction of $(\delta X_1, \delta X_2,
\delta Y_1, \delta Y_2) = (0, 0, -0.6546, 0.7560)$, corresponding
to a purely axion direction. This is the initial direction of the
slow roll away from the saddle point towards the nontrivial
minimum described above.

The value of the effective potential at the saddle point is $V \sim
3.35 \times 10^{{-16}} = M^{4}$ in Planck
units, so that the scale of inflation is $M =
3.25\times 10^{14}$ GeV. This is a rather small scale. The ratio of 
tensor to scalar perturbations in this scenario is very small, 
$r\ll 1$, so the gravitational waves produced in this scenario will be very
hard to observe. 

To find the slow roll parameter $\eta$ at the saddle point
(recall that $\epsilon = 0$ automatically at a saddle point), as well as to compute
the inflationary trajectories, we must use  the generalized
definitions of the slow-roll parameters.

This leads to
$n_{s} \approx 0.95$ and a long period of inflation, 980
$e$-foldings after the end of eternal inflation.

\begin{figure}
\centering
\includegraphics[height=5cm]{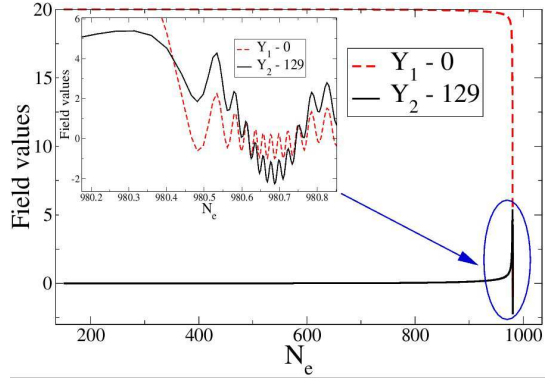}
\caption{Evolution of the axionic fields $Y_1,Y_2$ during inflation. We plot the values of $Y_i-Y_i^{min}$.
Inset shows oscillations around the minimum at the end of 
inflation.}
\label{end4.eps}       
\end{figure}

\begin{figure}
\centering
\includegraphics[height=5cm]{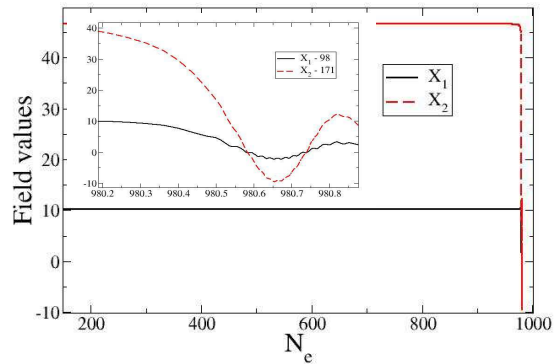}
\caption{ Evolution of the $X_1,X_2$ directions
during the inflationary period. We plot the values of $X_i-X_i^{min}$.
Inset shows oscillations around the minimum at the end of 
inflation.}
\label{x1x2.eps}       
\end{figure}

We have computed the power spectrum for the model under consideration
by first
numerically evolving the full set of field equations, which can
be efficiently written in the form
\bea
{d\phi_i\over dN} &=& {1\over H} {\dot\phi_i}(\pi_i) \nonumber\\
{d\pi_i\over dN} &=& -3\pi_i -{1\over H} {\partial\over
\partial\phi_i}\left(V(\phi_i) - {\cal L}_{\rm kin}\right)
\eea 
Here $
    {\cal L}_{\rm kin}$ is defined in Eq.   (\ref{kin})
and  $N$ is the number of $e$-foldings starting from the
beginning of inflation, $\pi_i =
\partial{\cal L}_{\rm kin}/\partial\dot\phi_i$ are the canonical
momenta, and the time derivatives ${\dot\phi_i}$ are regarded as
functions of $\pi_i$.  We use initial conditions where the field
starts from rest along the unstable direction, close
enough to the saddle point to give more than 60 e-foldings
of inflation. In fact our starting point corresponds to the
boundary of the eternally inflating region around the
saddle point.
  An example of the inflationary trajectories for all the 
fields is shown in figures \ref{end4.eps}, \ref{x1x2.eps}.
The choice of the satisfactory parameters for this model
is not unique, just as in the racetrack scenario
 \cite{Blanco-Pillado:2004ns}.  There is a rescaling of parameters
which does not alter inflationary dynamics or the
height of the potential; it rescales the fields, but leaves the
slow-roll parameters and the amplitude of density perturbations
invariant. There is also a second set of rescalings, which does rescale the potential and the amplitude of density fluctuations.

Evaluating the spectral index at $55$ $e$-foldings before the end
of inflation gives the spectral properties relevant for the CMB.
For $W_0 = 5.227\times 10^{-6}$ the
spectral index reaches its largest value 
\be
    n_s  \approx  0.95 \ .
\ee 
This is the same value that was found in the original
racetrack model. The value of  $W_0$ has to be tuned at
the level of a percent to keep the spectral index from decreasing
into a range of phenomenologically disfavored values.

In the KKLT model, with the superpotential containing only one
exponent for the volume modulus, one could not have inflation without
adding moving branes. In the original racetrack inflation scenario
 \cite{Blanco-Pillado:2004ns} it was possible to find the first working inflationary
model without adding any new branes to the KKLT vacuum stabilization
scenario. In the better racetrack model inflation is  achieved
in a theory with two moduli fields, without introducing
the standard racetrack potentials with two exponential terms for each
of them. This suggests that by increasing the number of
moduli fields and/or the number of non-perturbative contributions to the superpotential,  inflation may become  easier to achieve and with less fine-tuning.

\subsection{Gaugino-condensation induced racetrack models of inflation}

An interesting development of racetrack models of inflation was suggested in \cite{Lalak:2005hr}. They did not use the flux contribution to the superpotential, i. e. they choose $W_0=0$. They have assumed that the total volume modulus is fixed, and focused their attention on the dilaton-axion field and one more complex field with the canonical K\"ahler potential,
\be
K=-\ln (S+\bar S) +\chi \bar \chi \ .
\ee
The choice of the non-perturbative superpotential for these two fields was made on the basis of some prior studies of the effects of the gaugino condensation where the superpotential may have in addition to the exponential dependence on the dilaton-axion also a specific dependence on other moduli. In their particular model   
\be
W= \chi^p AN_1 M^3 e^{-S/N_1} +\chi^{p'}BN_2 M^3 e^{-S/N_2} \left({M^2\over (\alpha +\beta \chi)^2}\right)^\gamma \ .
\ee
Thus the model has two complex moduli, $S=s+i\phi$ and $\chi= xe^{i\theta}$, and   depends on 9 parameters. The specific choice was made for a good inflationary model:  $A=1.5, B=8.2$, $N_1=10, N_2=9$. The additional interactions/parameters, which were absent in previous models, are
$p=p'=0.5$, $\alpha=1, \beta=2.3$ and finally $\gamma=10^{-4}$. The model does not seem to require a significant fine-tuning. However, the smallness of $\gamma$ is important as the mass-squared of $\theta$ is proportional to $\gamma$. The qualitative picture of this model is very close to the racetrack models described above in a sense that the potential has a saddle point with the flat $\theta$-axion direction, where the eternal topological inflation may take place. Inflation ends with an exit stage, when the system reaches  the minimum of the potential and all  moduli are stabilized.

\subsection{Inflation in models with large volume of compactification}

Another interesting class on inflationary models \cite{Conlon:2005jm,Bond:2006nc} was developed on the basis of the so-called large volume compactification \cite{Balasubramanian:2005zx}, where the value of the $W_0$ in the superpotential is not small, as in simplest KKLT models, and therefore the $\alpha'$ correction to the K\"ahler potential play a significant role. An example of a successful model is codified in the K\"ahler potential
\be\label{Kahler2}
   K =  - 2 \ln\left[\alpha \left({(T_1+\bar T_1)}^{3/2} - \lambda_2{(T_2+\bar T_2) }^{3/2}- \lambda_3{(T_3+\bar T_3)}^{3/2} \right)+{\xi\over 2}\right]
   \,,
\ee
and the superpotential 
\be
W=W_0+ \sum_{i=1}^3 A_i e^{a_1 T_i} \ .
\ee
Here the term $\xi/2$ is due to the $\alpha'$ corrections to the K\"ahler potential. The model has 11 parameters, to compare with 5 parameters in the previously discussed racetrack and better racetrack models. It offers therefore more possibilities to look for inflationary slopes. The choice made in \cite{Conlon:2005jm}, as well as in roulette case \cite{Bond:2006nc}, was first of all to find the conditions on the parameters of the model which allow to stabilize the moduli $T_1$ and $T_2$ in a way that they actually do not participate in inflation; only the third modulus, $T_3$ is driving inflation. In \cite{Conlon:2005jm} the axion $\rm Im T_3$ of the third modulus $T_3$ is also frozen at its minimum, and the inflaton is given by $\rm Re T_3$. The evolution of this field takes place in a nice $\tau$-trough.

In a more general case considered in \cite{Bond:2006nc}, two complex moduli are  fixed, as before, but in the third modulus, $T_3=\tau +i\theta$, both the volume modulus $\tau$ and the  axion $\theta$ are not fixed. As the result, there are many inflationary trajectories in the landscape of this model. The potential is exponentially flat in the $\tau$-direction and has a periodic structure in  the $\theta$-direction. 

The cosmological evolution of the complex field $T_3$ was evaluated numerically using the ``SuperCosmology''  code  \cite{Kallosh:2004rs} designed for models with generic moduli space metric and arbitrary number of complex fields with any K\"ahler potential and superpotential. 
The trajectories depend on initial conditions in the $\tau, \theta$ plane.
The  randomness of $(\tau, \theta)$
initial conditions allows for a large ensemble of trajectories.
Features in the ensemble of histories include roulette trajectories,
with long-lasting inflation in the direction of the rolling axion,
enhanced in number of e-foldings over those restricted to lie in the
$\tau$-trough. Asymptotic flatness of the potential makes possible an
eternal slow-roll inflation. A wide variety of
potentials and inflaton trajectories agree with the cosmic microwave
background and large scale structure data. 

The A)B)C) for this class of models is simple, the value of the $n_s$ is predicted to be $0.96$ in   agreement with current data, no observable gravitational waves, no cosmic strings.

\section{  N-flation/assisted inflation}

At present N-flation \cite{Dimopoulos:2005ac,Easther:2005zr} is the  only  inflationary models studied in the context of string theory that may result in an effective chaotic inflation with a significant level of gravitational waves. It is close to the ideas of assisted inflation proposed earlier in \cite{Liddle:1998jc,Kanti:1999ie} The main idea of assisted inflation is that each field feels the downward force from its own potential but also the collective frictional force from all fields. Therefore slow-roll is  easier to achieve and the individual fields do not have to exceed  Planck scale vev's. 

The equations of motion for a set of scalar fields in generic situation  with the moduli space metric  (in real notation with $L_{kin}= {1\over 2} G_{ij}(\phi)\partial_\mu  \phi^i \partial^\mu  \phi^j $) are \cite{Gaillard:1995az,Sasaki:1995aw}: 
\be
\ddot \phi^i + \Gamma^i_{jk}(\phi) \dot \phi^j \dot \phi^k + 
3 H \phi^i +G^{ij}(\phi)\partial_j V=0 \ .
\label{modulispace} \ee
Here $G^{ij}(\phi)$ is the inverse metric of the moduli space and $\Gamma^i_{jk}(\phi)$ are the Christoffel symbols in the moduli space.
If  $G_{ij}=\delta_{ij}$ and $V=\sum_i V_i(\phi_i)$, i.e., if the metric of the moduli space is flat and if the potential is a sum of the potentials of the individual fields, the assistance effect becomes clear: 
\be
\ddot \phi^i + 
3 H \phi^i +\partial_i V_i=0\ ,   \qquad H^2= {\sum_i V_i\over 3 M_{Pl}^2} \ .
\label{flat}\ee
Each field responds to its own potential (there is no summation in the term $\partial_i V_i$ above), but the friction via the Hubble parameter comes from all fields and can be significantly stronger than in case without assistance.

Thus in general case of moduli space metric we need to identify in string theory the situations  when the simplified Eq.   (\ref{flat}) gives a good approximation for the actual complicated dynamics given by  Eq.   (\ref{modulispace}). 

An interesting attempt to do so was made in \cite{Dimopoulos:2005ac}.
The model requires a large number  of axions,  $
N\sim  240  \left ({M_{Pl}\over f}\right )^2
$,
where $f$ is the generic axion decay constant. For $f\approx 10^{-1}M_{PL}$, one should have $N\approx 10^4$. String theory may provide such a large number of axions, there are known examples of up to $10^5$ axions.   
If $f$ is a smaller fraction of $M_{Pl}$ the number of required axions grows. 

One may wonder whether the phenomenological assumptions made in \cite{Dimopoulos:2005ac,Easther:2005zr} can be justified in the known framework of compactified string theory. The main assumption  is that in the effective supergravity model with  numerous complex moduli, 
\be
t_n= {\phi_n\over f_n}+i M^2 R^2_n \ ,
\ee 
all moduli $R_n^2$ quickly go to their minima. Then only the axions ${\phi_n\over f_n}$ remain to drive inflation.  The reason for this assumption is that the K\"ahler potential depends only on the volume modulus of all two-cycles, $R_n^2 =-{i\over  2M^2} (t_n-\bar t_n)$, but is does not depend on the axions ${\phi_n\over f_n}= {1\over 2}( t_n+\bar t_n)$, so one could expect that the axion direction in the first approximation remains flat. Let us examine this assumption more carefully.

The K\"ahler potential is given by the same formula as the one in Eq.   (\ref{cubic}), which is an exact expression for ${\cal N}=2$ supergravities with cubic prepotentials.
 If superpotential does not depend on $t_n$, the potential  has a runaway dependence on the moduli $R_n^2$:
\be
V\sim e^K = {1\over {i C^{lmn}\over  3!}(t_l-\bar t_l) (t_m-\bar t_m) ( t_n- \bar t_n)}
\ee 
Here the Calabi-Yau intersection numbers $C^{lmn}$ can be positive as well as negative.  In this approximation, the potential   is  flat in the axion directions, but the vacuum  is unstable.

The instanton contribution to the superpotential, 
\be
W= \sum_n w_n e^{2\pi t_n}\ ,
\ee
 is supposed to stabilize the volume moduli $R_n^2$ quickly, whereas the axions are expected to slowly reach their minima. When all volume moduli are fixed, the moduli space metric is flat, up to some constant rescalings.
This assumption makes it easy to establish that the total potential with stabilized volume moduli is reduced to the {\it sum of potentials of each axion}, since the problem is reduced to the global SUSY potential
\be
V\approx  \sum_n V_n \approx \sum _n |\partial_n W|^2 \ .
\ee
Thus both conditions specified above, the flatness of the moduli space metric and the decoupling of the  potentials, are supposed to be satisfied \cite{Dimopoulos:2005ac,Easther:2005zr}.
This leads to the uncoupled set of light massive axions,  which makes the assistance effect easily possible in accordance  with Eq.   (\ref{flat}).

Can the assumption that {\it volume moduli stabilize quickly and axions slow roll for a long time} be justified in known models of string theory which stabilize both types of complex fields? 

The study of the potentials for $10^4$ fields is difficult to carry out. However, we have some experience with exactly the same type of KKLT potentials in case of one or two complex moduli, which we described in the previous sections of this paper. The generic property of this class of models is that the same exponential terms in the superpotential that stabilize  the volume moduli simultaneously stabilize the  axions. Therefore the  masses of these two fields are not much different. One can easily see it in the  simplest model  of KKLT potential,  Fig. 2.

We can now look at the more complicated examples. The racetrack potential has a complicated profile, shown in Figs.  4, 5. Near the saddle point the axion has a flat   maximum, the volume is at the local minimum. Therefore the significant  part of inflation proceeds via the slow-roll of the axion till the motion of the axion kicks the volume modulus out of the local minimum so that   
both fields reach the absolute minimum in the waterfall type evolution. However, one can easily see in Fig. 5 that  near the minimum there is no significant flatness in the axion direction comparative to the volume modulus direction, same as in the  simplest KKLT potential in Fig. 2. 

In the better racetrack model the potential near the saddle point for the two axions and another one for the two volume moduli are plotted in Fig. 6.  The trajectories for the two axions and the volume moduli  are shown in Figs. 7,8. Here again, near the saddle point there is a flat maximum in the  direction which is a combination of two axions. However, near the absolute minimum all directions are steep.

A related general observation was recently made in the models with large volume of compactification where  various trajectories with different initial condition were studied. It was stressed in \cite{Bond:2006nc} that it was easy to find the trajectories evolving only in the volume-$\tau$ direction, however, so far no inflationary trajectories in the axion-$\theta$ direction were found in this model. 

Thus all known string theory models in which the stabilization of the scalar and the pseudo-scalar field is possible due to the exponential terms in the superpotential do indeed have a $\cos $-type dependence on the axion field.  However, comparing with potentials in  (\ref{potential1}), (\ref{scalarpot2}),  one can see that
they are quite different from the simplified PNGB (Pseudo-Nambu-Goldstone Boson) potential $V=\Lambda^4[1- \cos(\phi/f)]$ with the fixed radial field.  In case of two axions $Y_1, Y_2$ and two  volume moduli $X_1, X_2$  in Eq.   (\ref{scalarpot2}) there is a term with $\cos(aY_1)$,  $\cos(bY_2)$, as well as a mixing term $\cos(bY_2- aY_1)$. More importantly, the dependence on $X_1, X_2$ is rather involved, which leads to a complicated dynamics, in general. 

It may be useful to compare the actual potential (\ref{scalarpot2})  with the simplified potential depending on  two axions in \cite{Kim:2004rp}, where it was argued that the models with two axions may lead to an assistance effect. However, the main assumption in \cite{Kim:2004rp} is that the potential of the form $V=\Lambda_1^4[1- \cos(a\phi_1)]+
\Lambda_2^4[1- \cos(b \phi_1+ c \phi_2)] $ can be derived from string theory with the superpotential  (\ref{dosexp}). This assumption is problematic at this stage because of the volume stabilization issue. Both axions and volume moduli undergo a dynamical evolution according to the potentials in eqs. (\ref{potential1}), (\ref{scalarpot2}) and  the use of the pure multi-axion potential is not valid for the known models derived from string theory. 

The axion valley model proposed in Sec. 3.1 of this paper, see Fig. 3,   would support the ideas in \cite{Kim:2004rp},  and it is valid in supergravity.  But this model is still to be derived from string theory.

\

It may still be possible to justify the N-flation model in string theory. For example, one can try to find a string theory realization of the  axion valley  model described in Sec. 3.1, or of some other model of this kind where the basic assumptions made in  \cite{Dimopoulos:2005ac,Easther:2005zr} are satisfied.

Another possibility is to start with initial conditions where all volume moduli are very close to the minimum of the potential and axions are away from the minimum. If the motion of all axions leads to a significant friction, because of the assisted inflation, one may hope that the volume moduli will stay near their minima, and the regime of N-flation model will be valid. This is not a very attractive proposition since it requires a lot of additional fine-tuning of initial conditions.  

Finally, it may happen that even if one considers a simultaneous motion of all interacting fields, the axions and the radial moduli, the simple fact that there are many of them may be sufficient for the existence of the assisted inflation regime. This possibility requires a more detailed investigation.

Thus, the assisted models of inflation in string theory require more work, just as all other models described in this talk. Assisted  inflation in string theory will become a particularly important issue if the gravitational waves from inflation will be detected.

\section{Discussion}

String cosmology has several different but closely related goals: to find inflationary models based on string theory, to identify some of their predictions which may be related to the specifically stringy nature of the inflationary models, and, by doing so, to test string theory by comparing its predictions with observations.  

In discussion of inflationary models in string theory and supergravity we  described shift symmetry which may under certain conditions explain the required flatness of the inflaton direction of the potential. Examples include   chaotic inflation in supergravity model \cite{Kawasaki:2000yn} and the axion valley model (supergravity version of the natural inflation) proposed here in Sec. 3.1. In the axion valley model shown in Fig. 3 the scalar is heavy and quickly stabilizes, whereas the pseudo-scalar remains very light before it reaches the minimum of the potential. This is in contrast with the KKLT model where both the scalar and the pseudo-scalar (the volume modulus and the axion) have approximately the same curvature of the potential near the minimum, as shown in Fig. 2.  The supergravity version of chaotic inflation and the axion valley model  use the shift symmetry of the K\"ahler potential in a very effective way. Therefore it would be most interesting   to find string theory versions of these models.

An inflationary D3/D7 brane model \cite{Dasgupta:2002ew,Koyama:2003yc,Dasgupta:2004dw} discussed in Sec. 4.2 is also based on  shift symmetry 
slightly broken by quantum corrections \cite{Hsu:2003cy,Firouzjahi:2003zy,Hsu:2004hi}. 
In D3/D7 brane model   the shift symmetry  originates from  the ${\cal N}=2$ supergravity structure of the K\"ahler potential shown in Eqs.   (\ref{N2}), (\ref{cubic}). In ${\cal N}=2$ models the K\"ahler potential \footnote{It is interesting that the theory of ${\cal N}=2$ supersymmetric black holes in string theory and the attractor mechanism stabilizing the moduli  near the black hole horizon are both based on precisely the same class of K\"ahler potentials, see for example Eq.   (30) in \cite{Ferrara:1995ih}.}  is given by $K=-\ln \left [i {C_{ABC}\over 3!}(z-\bar z)^A (z-\bar z)^B (z-\bar z)^C\right ]$ and the shift symmetry
is generic. Here  $C_{ABC}$ are the intersection numbers of the compactification manifold. The ${\cal N}=2$ supersymmetry of these models may be broken by fluxes and/or non-perturbative corrections, but the total potential may still enjoy a slightly broken  shift symmetry under the transformation of the inflaton field $z\rightarrow z+ \delta$, with real parameter $ \delta$. In the racetrack models in Sec. 5.1 and 5.2 the K\"ahler potentials also have  shift symmetry, since they are given by  the  ${\cal N}=2$ formula, see for example Eq. (\ref{Kahler}). The shift symmetry of the total potential is, however,  strongly broken near the minimum and slightly broken near the saddle point.

In other models which we discussed here the shift symmetry was not relevant for the flatness of the inflaton potential. It was  not present in  the KKLMMT model \cite{Kachru:2003sx}, which therefore in most cases relies on a special dependence on the inflaton field in the superpotential to cancel the one in the K\"ahler potential. Such a cancellation requires fine-tuning, which may be possible in the context of a huge stringy landscape.

In the inflationary models with extra large  volume of compactification \cite{Conlon:2005jm,Bond:2006nc} the dynamics was mostly based on the very flat radial modulus direction. There is no shift symmetry of the K\"ahler potential and of  the total scalar potential in the inflaton direction $T_3+\bar T_3$ on which the K\"ahler potential depends strongly. The flatness is achieved via some set of hierarchy relations between the parameters of the model with three complex fields. The flatness in the axion direction (the remnant of the shift symmetry of the K\"ahler potential in Eq. (\ref{Kahler2})) does help to increase the number of e-foldings \cite{Bond:2006nc}, but it is not a major feature of the model.

We also discussed the string theory  N-flation model \cite{Dimopoulos:2005ac,Easther:2005zr} of assisted inflation \cite{Liddle:1998jc,Kanti:1999ie}.   We pointed out that the  assumption in \cite{Dimopoulos:2005ac,Easther:2005zr} that only the axions define the dynamics due to the shift symmetry of the K\"ahler potential may be valid in the context of the axion valley supergravity model presented in  Sec. 3.1 here, but not in the currently known constructions of string theory.  We explained, in particular,  why this assumption is not valid  in the simplest KKLT model of moduli stabilization. 

Nevertheless, it might be possible to find an assisted inflation regime in string theory  by investigation of a combined dynamics of all string theory moduli. This, as well as finding a string theory generalization of the supergravity versions of chaotic and natural inflation discussed above, will be particularly important if the primordial gravitational waves from inflation will be detected. 

In this paper we described a representative subclass of string inflation models, but nevertheless the list of the models which we discussed is certainly incomplete. Moreover, one should clearly understand that the whole subject is relatively new, and it is difficult to make any far-reaching conclusions at this stage. For example, the time span between the invention of the simplest chaotic inflation model $m^{2}\phi^{2}$ \cite{Linde:1983gd} and its supergravity version \cite{Kawasaki:2000yn} was about 17 years. Similarly, the time span between the invention of the natural inflation \cite{Freese:1990rb} and its implementation in supergravity (see Sect. 3.1 of this paper) was also about 17 years.  
Meanwhile the KKLT mechanism of vacuum stabilization was proposed only 4 years ago, and therefore the development of the cosmological models based on string theory with a stabilized vacuum began only very recently.

The search of new models of string theory inflation should go in parallel with the further development of string theory. One may try to find some new theoretical structures  which will lead to interesting inflationary models in string theory. This is  one of the important challenges for string theory and cosmology. String theory community is well aware of these challenges, see for example, \cite{Douglas:2006es, Blumenhagen:2006ci,Ooguri:2006in,BBS}, and one may hope that a better understanding of the core string theory may lead to  better models for cosmology.

Under certain conditions, the new developments may allow us to test string theory by current and future precision data in cosmology. The conditions include a reliable derivation of  inflationary models from string theory. These models should have distinct predictions for observables like the spectral index $n_s$, the level of gravitational waves, and the abundance of light cosmic strings.
If these conditions are satisfied,  one may hope that in few years from now, when the precision data will become available and the derivation of the inflationary models will be refined,  it will be possible to test the string theory  assumptions underlying the derivation of the  corresponding inflationary models.

\section{Acknowledgements}

I am grateful to the organizers and participants of ``Inflation+25'' for the most stimulating atmosphere of the conference. I thank all  my collaborators on the projects described in the talk and     L. McAllister, T. Banks, D. Baumann, R. Bean, S. Dimopoulos, S. Kachru, L. Kofman, A. Linde, D. Lyth, L. McAllister,  E. Silverstein, H. Tye  and J. Wacker for the discussions of various topics in this talk. This work was supported by the NSF grant 0244728 and by the A. von Humboldt award.

\bibliographystyle{toine}

 \bibliography{InflRepl}



\printindex
\end{document}